\documentclass[article,nofootinbib]{revtex4-1}
\usepackage{fullpage,amsmath,physics,hyperref}
\usepackage{graphicx}
\usepackage[english]{babel,xcolor}
\usepackage{amssymb,mathtools}
\usepackage{braket}
\usepackage{subfig}

\newcommand{\figcontour}{
\begin{figure}
        \centering
        \includegraphics[scale=0.5]{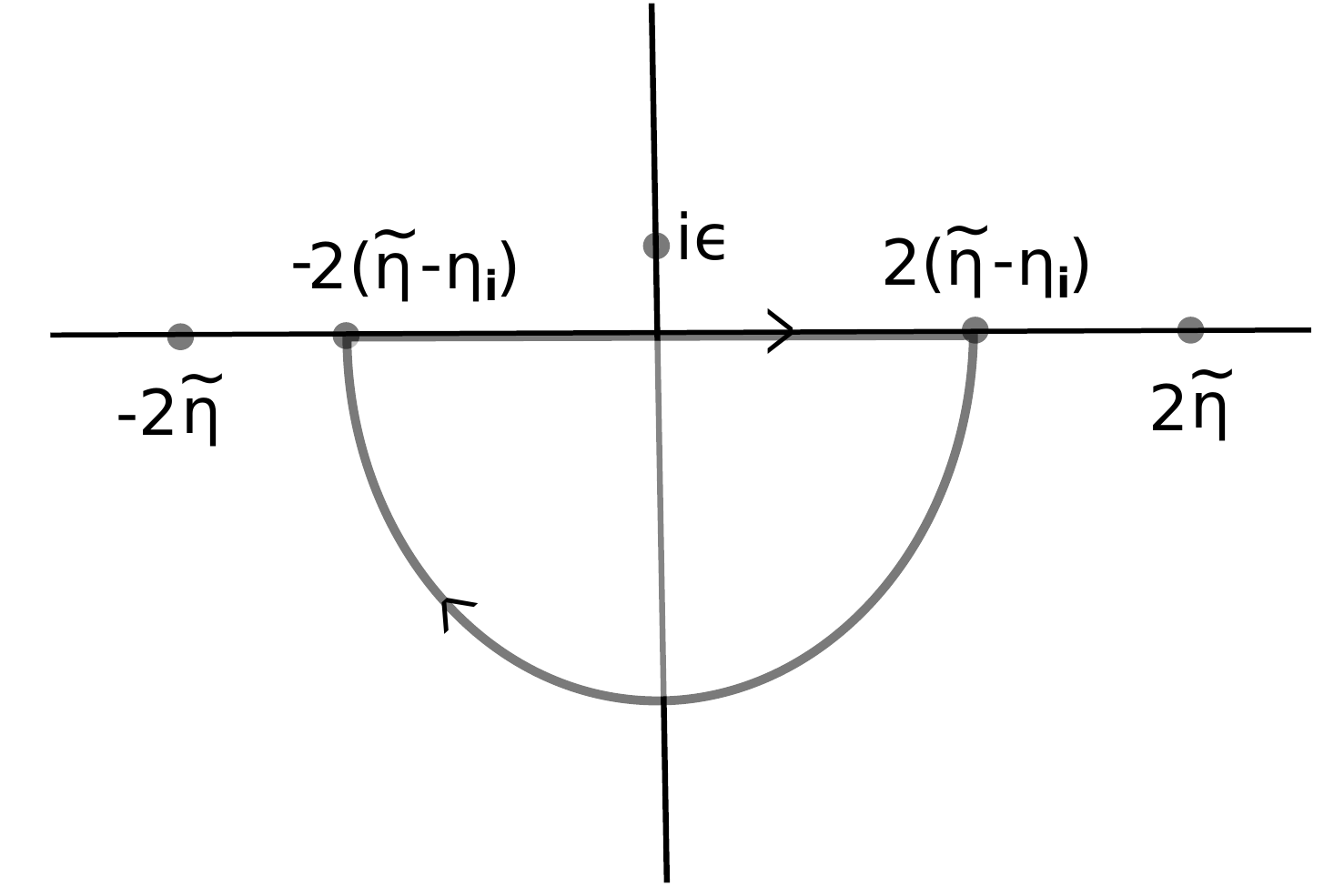}           
        \caption{The chosen contour does not contain the poles inside it.  }\label{cont}
    \end{figure}}

\begin{document}
\title{Unruh deWitt probe of late time revival of quantum correlations in Friedmann spacetimes}
\author{Ankit Dhanuka}
\email{ankitdhanuka555@gmail.com, ph17006@iisermohali.ac.in}

\author{Kinjalk Lochan}
\email{kinjalk@iisermohali.ac.in}
\affiliation{Department of Physical Sciences, IISER Mohali, \\ Sector 81, SAS Nagar, Manauli PO 140306, Punjab, India}
\begin{abstract}
\noindent Unruh deWitt detectors  are important constructs in studying the dynamics of quantum fields in any geometric background. Curvature also plays an important  role in setting up the correlations of a quantum field in a given spacetime.  For instance, massless fields are known to have large correlations in de Sitter space as well as in certain class of Friedmann-Robertson-Walker (FRW) universes. However, some of the correlations are secular in nature while some are dynamic and spacetime dependent. An Unruh deWitt detector responds to such divergences differently in different spacetimes.
In this work, we study the response rate of Unruh deWitt detectors which interact with quantum fields in FRW spacetimes. We consider both conventionally as well as derivatively coupled Unruh deWitt detectors. Particularly, we consider their interaction with massless scalar fields in FRW spacetimes and nearly massless scalar fields in de Sitter spacetime. We discuss how the term which gives rise to the infrared divergence in the massless limit in de Sitter spacetime manifests itself at the level of the response rate of these Unruh deWitt detectors in a wide class of Friedmann spacetimes. In order to carry out this study, we make use of an equivalence that exists between massless scalar fields in FRW spacetimes with massive scalar fields in de Sitter spacetime. Further, we show that while the derivative coupling regulates the divergence appearing in de Sitter spacetime, it does not completely remove them in matter dominated universe. This gives rise to large transitions in the detector which can be used as  a probe of setting up of large correlations in late time era of the universe as well. We also apply the results of these otherwise formal analyses to the coupling of hydrogen atoms with gravitational waves. We show that the coupling of hydrogen atoms with gravitational waves takes a form that is similar to derivatively coupled UdW detectors and hence has significant observational implications as a probe of late time revival of quantum correlators.   
\end{abstract}
\maketitle

\section{Introduction}

\noindent Unruh deWitt (UdW) detectors are quantum probes which follow classical trajectories in spacetime while measuring effects of quantum correlations of a background field on quantum systems \cite{Birrell:1982ix}. Other than their classical motion in spacetime, they have an internal quantum structure with discrete quantum levels. These detectors couple to quantum fields and the coupling of a detector with a quantum field can cause the detector to make transitions between its internal quantum levels. The probability amplitude for a detector to undergo these transitions depends crucially upon the state of the quantum field and the trajectory of the detector in spacetime. The coupling of particle detectors with quantum fields (and hence the probability amplitude) senses the correlations of quantum field in the state in which it is placed and the response of the detector along any particular trajectory also encapsulates in it the fact that the particle content of a quantum field in any state is an observer dependent quantity \cite{Unruh:1976db, Crispino:2007eb}. Different particle detectors differ from each other by their internal quantum structure (i.e., whether it is a two-level system \cite{Crispino:2007eb, Svaiter:1992xt}, a quantum harmonic oscillator \cite{Brown:2012pw,Lin:2006jw}, etc.), their interaction with the field (i.e., whether it employs a monopole coupling, a derivative coupling \cite{Juarez-Aubry:2014jba, Louko:2014aba, Martin-Martinez:2014qda, Tjoa:2022oxv,Ford:1993bw,DePaola:1996xw}, etc.) and other things like whether they are operative for a finite time \cite{Sriramkumar:1994pb} or indefinite time or has some other form for the switching function \cite{Louko:2007mu} or whether they are taken to be point sized or they have some finite spatial size \cite{Schlicht:2003iy,Louko:2006zv,Lee:2012tvc,Martin-Martinez:2020pss}, etc. Though particle detectors have been traditionally used to study quantum field theory  content in non-inertial settings or  classical gravitational settings \cite{Unruh:1976db,Crispino:2007eb}, etc,  they have also been employed to investigate quantum effects of gravity on sensitive observables such as the entanglement between two entangled UdW detectors with different types of relative motion between them \cite{Lin:2008jj,Lin:2008yz} (for more on observer dependent entanglement, refer to \cite{Alsing:2012wf} and references therein).  
\\
Interaction of electromagnetic waves with atoms can be modelled by UdW type coupling \cite{Alhambra:2013uja, Martin-Martinez:2012ysv} and hence quantum optical setups can be used to test the predictions obtained by analysing UdW type couplings. For example, one can try to test the prediction that detectors moving along inertial trajectories in flat spacetime have a vanishing response rate whereas a UdW detector sees a thermal response rate when it follows a uniformly accelerating trajectory in flat spacetime \cite{Unruh:1976db, Crispino:2007eb}. 
In addition to the investigation of UdW detectors in flat spacetime, it is also important as well as interesting to analyse how curvature contributes to the response rate of UdW detectors in curved spacetimes \cite{K:2021gns}. One natural arena where curvature is present in the analysis is that of the evolution of our own universe where different epochs can be approximated by Friedmann-Robertson-Walker (FRW) spacetimes. In fact, the quantum dynamics of metric fluctuations over these spatially homogeneous and isotropic background FRW spacetimes play an important role in shaping the present day universe the way we observe it \cite{Weinberg:2008zzc}. Quantum field theories in FRW spacetimes have been studied extensively \cite{ Parker:1968mv, Parker:1969au, Parker:1971pt, Birrell:1982ix, Brandenberger:1984cz, Parker:1999td, Parker:2009uva, Lochan:2018pzs}  and they provide important insights into the formal aspects of QFT as well as into our universe.  A number of past studies, \cite{Garbrecht:2004du,Garbrecht:2004ui,Chakraborty:2019ltu,Hotta:2020pmq,Ali:2020gij, Conroy:2022rgp}, have also analysed the behaviour of quantum fields in FRW spacetimes by coupling them to UdW type particle detectors. For example, \cite{Garbrecht:2004du} studies the response rate of UdW detector which are coupled with real quantum scalar fields in the conformal as well as massless case in de Sitter spacetime, \cite{Ali:2020gij} studies the response rate for quadratically coupled complex scalar fields again in both conformal as well as massless cases in de Sitter spacetime, \cite{Chakraborty:2019ltu} studies the transition probability for conformal vacua in FRW spacetimes. One important and well known feature of quantum fields in a class of FRW spacetimes is that their Wightman functions have infrared divergences \cite{Ford:1977in,Allen:1985ux,Antoniadis:1985pj,Allen:1987tz,Polarski:1991ek,Miao:2010vs,Lochan:2018pzs}. Such a divergence has a root in the fact that massless fields in power law FRW universes are conformally equivalent to massful fields in de Sitter spacetime, which may harbour such divergences in correlations. It has been argued recently \cite{Stargen:2021vtg} that potential divergences in correlation functions strongly enhance the UdW responses to reveal small acceleration dependence. Therefore, we expect that the infrared divergences in FRW spacetimes should also lead to enhancement of the UdW response rates. With this motivation, we seek to investigate the coupling of UdW detectors with quantum fields in FRW spacetimes.  \\
\noindent First, we consider the conventional Unruh deWitt (UdW) type coupling where there is a monopole coupling between the field and the operator which causes the transitions in the internal quantum space of the detector \cite{Crispino:2007eb}. In this case, the response rate of transition between quantum states of the detector is related to the Wightman function of the quantum field in the considered spacetime. Therefore, the behaviour of the correlations of quantum field between spacetime points along the trajectory of the detector is imprinted in the expression of the transition response rate. In the case of de Sitter spacetime, we consider nearly massless scalar fields and place them in the Bunch Davies vacuum \cite{Allen:1985ux}. The Wightman function of a scalar field in de Sitter spacetime has an infrared divergence \cite{Allen:1987tz,Polarski:1991ek,Kirsten:1993ug,Miao:2010vs} in the mass going to zero limit. The term corresponding to the infrared divergence has no spacetime dependence and it provides a dominant secular contribution to the response rate of the detector. \\ 
We also consider massless scalar fields in radiation dominated spacetime and matter dominated spacetimes to study detector response in other epochs of cosmological expansion. For these cases, we make use of an equivalence between massless scalar fields in FRW spacetimes with that of massive scalar fields in de Sitter spacetime \cite{Lochan:2018pzs,Dhanuka:2020yxp,Lochan:2022dht}. Using this equivalence, we place massless scalar fields in FRW spacetimes in the Bunch Davies like vacuum of the corresponding massive scalar field in de Sitter spacetime. The Wightman function of massless fields in matter dominated cases inherit the infrared divergence of the de Sitter spacetime but now with a time dependent conformal factor multiplying the divergent term \cite{Lochan:2018pzs,Dhanuka:2020yxp}. Thus, we find that the term corresponding to the infrared divergence provides the dominant contribution to the transition response rate. For radiation dominated case, the massless scalar field correlator does not possess any such infrared divergent term and hence provides finite detector response. \\
\noindent The analysis shows that the infrared divergences of the de Sitter and matter dominated spacetimes manifest themselves in the detector response.  However, in de Sitter spacetime the divergence of correlators is sometimes argued to be originated from breaking of de Sitter symmetry \cite{Allen:1985ux, Allen:1987tz} and any physically sensible result should be free of any divergences. A line of argument to that end is to regard only those operators as physical which are infrared finite. For example, \cite{Page:2012fn} argues that the shift invariant operators like the differences of the field operators, derivatives of fields etc., are to be regarded as true physical observables as they are infrared finite for massless scalar fields in de Sitter spacetime. Similarly, the derivatives present in the stress energy operator also renders it infrared finite for the de Sitter spacetime \cite{Ford:1984hs}. Keeping these arguments in mind, we look at the response of more 'physical' derivatively coupled UdW detectors.   \\
In the derivative coupling case, the detector couples to the derivative of the field with respect to the proper time along the trajectory \cite{Juarez-Aubry:2014jba}. For this case, the response rate of transition between quantum states of the detector depends upon the double derivative of the Wightman function of the field with respect to the proper time at different points along the detector's trajectory \cite{Juarez-Aubry:2014jba, Tjoa:2020eqh}. In the case of de Sitter spacetime,  under the action of the derivatives, this term goes away as the infrared divergent term in the Wightman function for nearly massless scalar fields does not have any spacetime dependence. Thus the transition response rate of derivatively coupled UdW detector for nearly massless scalar fields in de Sitter spacetime remains finite. However, in the case of massless scalar fields in matter dominated spacetimes, the infrared divergent term has time dependence and it does not vanish under the action of derivatives with respect to the detector's proper time and provides the dominant contribution to the response rate. Thus, even though the derivative coupling could cure the infrared divergence of the de Sitter spacetime, this does not happen for the matter dominated spacetimes.  Using this we argue that the realistic physical systems, e.g. the derivatively couple UdW detectors are expected to capture the revival of quantum correlations in the matter dominaated era of the universe.  \\
In addition to these formal analyses of UdW detectors and derivatively coupled UdW detectors for quantum scalar fields in the considered spacetimes, we investigate the scenario of the coupling of hydrogen atoms with gravitational waves where a derivatively coupled UdW like coupling occurs. Following the treatment given in \cite{Parker:1980kw, Parker:1980hlc}, we consider the interaction of a non-relativistic hydrogen atom (whose center of mass is moving along some time-like classical trajectory) with the curvature of the spacetime. Considering gravitational wave perturbations over the homogeneous and isotropic FRW  backgrounds, one can find the form of the above interaction term upto to leading order in gravitational perturbations. The interaction between gravitational waves and hydrogen atom has the form of a generalized derivatively coupled UdW detector. For this setting, the above analysis of the response rate of derivatively coupled UdW detector in matter dominated spacetimes can be carried over and the implications of the dominant infrared term on the transition of the electron of the atom between its different atomic states can be investigated. Such an analysis also provides a potential avenue to look for observational signatures of quantized gravitational waves. \\
The rest of the paper is divided in four sections. In section \ref{UdW}, we consider conventional UdW detectors for scalar fields in de Sitter, radiation-dominated and matter-dominated spacetimes and calculate the response rate for them. In section \ref{dUdW},  we perform a similar analysis as is done in section \ref{UdW} but for derivatively coupled UdW detectors. In section \ref{Impli}, we consider a specific UdW coupling where the detector couples with the stress energy tensor of the field. In this section, we also look at the dynamics of a hydrogen atom in FRW spacetimes with and without gravitational wave perturbations which harbours a derivatively  coupled UdW detector like structure. In section \ref{con}, we summarize the results obtained in this paper and their implications.   
\section{Conventional UdW detectors}\label{UdW}
\noindent In this section, we consider the response rate of a conventional Unruh deWitt detector which couples with massless scalar fields in FRW spacetimes. A conventional Unruh deWitt detector couples with a quantum field by the following type of interaction term \cite{Birrell:1982ix,Crispino:2007eb} 
\begin{equation}\label{501}
H_{int} = c\hat{\mu}(\tau)\chi(\tau)\hat{\phi}(x(\tau)) \, ,
\end{equation}
where $\hat{\mu}(\tau)$ is the detector term which governs the transitions within the internal quantum structure of the detector and $\hat{\phi}(x(\tau))$ is a quantum field which the detector is coupled to. Here $\tau$ represents the proper time of the detector along its classical timelike trajectory, $x(\tau)$, and $\chi(\tau)$ is a real-valued switching function which decides how the detector is turned on and off \cite{Sriramkumar:1994pb, Fewster:2016ewy} . \\ \\
Let us consider the case when detector makes a transition from some state $\ket{0}_{D}$ to another state $\ket{\Omega}_{D}$ which have energies $0$ and $\Omega$, respectively and the field starts in some state, $\ket{\psi}$, while it is allowed to go to any final state which are traced over. The transition probability for this case, in first order time-dependent perturbation theory \cite{Birrell:1982ix}, is given by 
\begin{equation}\label{502}
    P_{0\rightarrow \Omega}=c^2| {}_{D}\braket{\Omega|\hat{\mu}(0)|0}_{D}|^2\iint d\tau_{1} d\tau_{2}\chi(\tau_{1})\chi(\tau_{2})e^{-i\Omega(\tau_{1}-\tau_{2})}G(x(\tau_{1}),x(\tau_{2})),
\end{equation}
    where $G(x(\tau_{1}),x(\tau_{2}))=\braket{\psi|\hat{\phi}(x(\tau_{1}))\hat{\phi}(x(\tau_{2}))|\psi}$ is the two point function of the quantum field in the state $\ket{\psi}$. \\ \\
\noindent For a switching function operating uniformly over $\tau_{i}$ to $\tau_{f}$, the transition probability is given by 
\begin{equation}\label{503}
    P_{0\rightarrow \Omega}=c^2|{}_{D}\braket{\Omega|\hat{\mu}(0)|0}_{D}|^2\int_{\tau_{i}}^{\tau_{f}} \int_{\tau_{i}}^{\tau_{f}} d\tau_{1} d\tau_{2}e^{-i\Omega(\tau_{1}-\tau_{2})}G(x(\tau_{1}),x(\tau_{2})) \, .
\end{equation}
As motivated above that we want to investigate the role of the curvature of cosmological space and the divergent structure of correlations of quantum fields in a class of these spacetimes  on the response rate of UdW detectors, we specialize to Friedmann spacetimes. We consider the case in which the UdW detector moves along comoving trajectories for which the spatial coordinates are fixed and the comoving time is the proper time. Thus, we can go to the conformal coordinates in which $d\tau = a(\eta)d\eta$ where $a(\eta)$ denotes the scaling factor of the FRW spacetime under consideration i.e., $ds^2 = a^{2}(\eta)(-d\eta^2 + d\vec{x}^2)$. The probability amplitude, expressed in conformal coordinates, is given by 
\begin{equation}\label{504}
    P_{0\rightarrow \Omega}=c^2|{}_{D}\braket{\Omega|\hat{\mu}(0)|0}_{D}|^2\int_{\eta_{i}}^{\eta_{f}} \int_{\eta_{i}}^{\eta_{f}} d\eta_{1} d\eta_{2}e^{-i\Omega(\tau(\eta_{1})-\tau(\eta_{2}))}a(\eta_{1})a(\eta_{2})G(x(\eta_{1}),x(\eta_{2})) \, ,
\end{equation}
where $\eta_{i}$ and $\eta_{f}$ are the values for the conformal coordinate corresponding to the $\tau_{i}$ and $\tau_{f}$, respectively.\\ 
We now make a change of variables and introduce new coordinates 
$\tilde{\eta} \equiv (\eta_{1} +\eta_{2})/2\ $ and $\Delta \eta \equiv \eta_{1} -\eta_{2}.$
For any fixed  $\tilde{\eta} \in (\eta_{i},(\eta_{i}+\eta_{f})/2)$, we have $ \eta_{2} \in (\eta_{i}, 2\tilde{\eta}-\eta_{i})$  and   $\Delta \eta \in (-2(\tilde{\eta}-\eta_{i}),2(\tilde{\eta}-\eta_{i})).$ 
Similarly, for any fixed value of $\tilde{\eta} \in ((\eta_{i}+\eta_{f})/2,\eta_{f})$, we have $\eta_{2} \in ( 2\tilde{\eta}-\eta_{f},\eta_{f})$ and $\Delta \eta \in (-2(\eta_{f}-\tilde{\eta}),2(\eta_{f}-\tilde{\eta})).$ 

Using Eq.~\eqref{504} and going to the $(\tilde{\eta}, \Delta \eta)$ coordinates, we find that the rate of transition probability with respect to $\tilde{\eta}$, for $\tilde{\eta} \in (\eta_{i},(\eta_{i}+\eta_{f})/2))$, is given by \\
\begin{multline}\label{506}
\frac{1}{c^2|{}_{D}\braket{\Omega|\hat{\mu}(0)|0}_{D}|^2}\frac{dP_{0\rightarrow \Omega}}{d\tilde{\eta}} = \int_{-2(\tilde{\eta} - \eta_{i})}^{2(\tilde{\eta} - \eta_{i})} d(\Delta\eta)\ e^{-i\Omega(\tau(\tilde{\eta} + (\Delta \eta)/2)-\tau(\tilde{\eta} - (\Delta \eta)/2))}\\ G\Big(x\big(\tilde{\eta} + (\Delta \eta)/2\big),x\big(\tilde{\eta} - (\Delta \eta)/2\big)\Big) a\big(\tilde{\eta} + (\Delta \eta)/2\big)a\big(\tilde{\eta} - (\Delta \eta)/2\big) \, ,
\end{multline}
whereas for $\tilde{\eta} \in ((\eta_{i}+\eta_{f})/2, \eta_{f})$, the rate of transition probability is given by \\
\begin{multline}\label{507}
\frac{1}{c^2|{}_{D}\braket{\Omega|\hat{\mu}(0)|0}_{D}|^2}\frac{dP_{0\rightarrow \Omega}}{d\tilde{\eta}} = \int_{-2( \eta_{f}-\tilde{\eta} )}^{2(\eta_{f} - \tilde{\eta} )} d(\Delta\eta)\ e^{-i\Omega(\tau(\tilde{\eta} + (\Delta \eta)/2)-\tau(\tilde{\eta} - (\Delta \eta)/2))}\\ G\Big(x\big(\tilde{\eta} + (\Delta \eta)/2\big),x\big(\tilde{\eta} - (\Delta \eta)/2\big)\Big) a\big(\tilde{\eta} + (\Delta \eta)/2\big)a\big(\tilde{\eta} - (\Delta \eta)/2\big) \, .
\end{multline}\\

\noindent 
In order to analyse the case of interest i.e., massless scalar fields in power-law type FRW spacetimes, we make use of an equivalence \cite{Lochan:2018pzs, Dhanuka:2020yxp, Lochan:2022dht} according to which a massless scalar field in an FRW spacetime with scaling factor, $a(\eta) = (H\eta)^{-q}$, can be mapped to a massive scalar field in de Sitter spacetime with $m^2 = H^2(1-q)(2+q)$.
One also finds that the Wightman functions in the two equivalent settings are related by the following relation
\begin{eqnarray}\label{510}
G^{FRW}(x_{1},x_{2}) = (H\eta_{1})^{q-1}(H\eta_{2})^{q-1}G^{dS}(x_{1},x_{2}) \, ,
\end{eqnarray}
 for more details, refer Appendix A.2 of \cite{Lochan:2018pzs}.\\ \\
As for the state in the corresponding de Sitter spacetime is concerned, we take that to be the Bunch-Davies vacuum \cite{Allen:1985ux} for which the Wightman function\footnote{Though we are considering Bunch Davies vacuum here, one could also consider other physically well-behaved normalizable states \cite{Lochan:2014xja}. Such well-behaved states also share the divergent Bunch Davies correlator structure in addition to their own characteristic features which, however, do not significantly alter the characteristics  used in our analysis \cite{Kundu:2011sg,Lochan:2014xja, Lochan:2022dht}.} is given by 
\begin{equation}\label{511}
G^{dS}(x_{1},x_{2}) = \frac{H^2}{16\pi^2}\Gamma\Big(\frac{3}{2} + \nu\Big)\Gamma\Big(\frac{3}{2} - \nu\Big) {}_2F_1\Big(\frac{3}{2} + \nu,\frac{3}{2} - \nu,2,1-\frac{y}{4}\Big) \, ,
\end{equation} 
where 
\begin{equation}\label{512}
y(x_{1},x_{2}) = \frac{-(\eta_{1}-\eta_{2} - i\epsilon)^2 + (\vec{x}_{1}-\vec{x}_{2})^2}{\eta_{1}\eta_{2}} \,  ,
\end{equation}
and $\nu = \sqrt{\frac{9}{4}-\frac{m^2}{H^2}}.$\\ 

\noindent From the formula that, $m^2 = H^2(1-q)(2+q)$, we see that the square of the mass is positive only for the cases in which $q \in [-2,1)$. We consider only those FRW spacetimes which have $q$ belonging to this range. Since $a(\eta) = (H\eta)^{-q}$,  we see that $\eta \in (0,\infty)$ corresponds to expanding spacetimes for $q \in [-2,0]$, whereas for $q \in [0,1)$, $\eta \in (-\infty, 0)$ corresponds to expanding spacetimes. Let us briefly look at the response rate for UdW detectors which remain operative for the full time range of these spacetimes. 
We argue in Appendix \ref{infres} that the infinite time response rate with respect to $\tilde{\eta}$ has the following dependence on $\Omega$ and $H$
\begin{equation}
\frac{1}{c^2|{}_{D}\braket{\Omega|\hat{\mu}(0)|0}_{D}|^2}\frac{dP_{0\rightarrow \Omega}}{d\tilde{\eta}} \propto (\Omega H^{-q})^{\frac{1}{1-q}} \, .
\end{equation} 
From this expression, we see that, for $q \in (-2,0)$, the exponent of $H$ is positive and hence the response rate increases with increasing $H$. While for $q \in (0,1)$, the exponent of $H$ is negative and the response rate decreases with increasing $H$. In fact, the Ricci scalar for FRW spacetimes is given by $R \propto H^{2q}$ and we conclude that, for $q \in (-2,0)$, it increases with decreasing $H$ while, for $q \in (0,1)$, it increases with increasing $H$. Hence, the behaviour of the response rate and the Ricci scalar with respect to $H$ (for an FRW spacetime) are opposite of each other.  Other important point to notice is that the response rate gets most enhanced with $H (>1)$ for $q = -2$ whereas for $H(<1)$, it is for $q=1$ case that the response rate is most significantly enhanced with $H$. As we see below that the infrared divergent factors in de Sitter and matter dominated spacetimes cause very fast transitions, the above conclusion implies that for these cases, the $H$ dependence also contributes maximally to the response rate compared to the other considered spacetimes.  

\noindent In standard cosmological setting universe remains in  a given phase only for a finite time, thus it will be worthwhile to consider the finite time response rates for different cosmological eras as we do next. 
\subsection{Nearly Massless Scalar Fields in de Sitter Spacetime }
\noindent First, let us look at the behaviour of the response rate of UdW detectors which couple with nearly massless scalar fields in de Sitter spacetime. By nearly massless scalar fields, we mean that $\nu = 3/2 - \delta$ where $\delta<<1$. Since for scalar fields in de Sitter spacetime, the Wightman function is known to suffer from infrared divergences in the mass going to zero limit \citep{Allen:1987tz,Polarski:1991ek,Kirsten:1993ug,Miao:2010vs}, we expect the response rate (which depends upon the Wightman function) to also show similar infrared divergence. 
To be more precise, we note that the Wightmann function, as a power series in mass $\delta$, is given by \cite{Lochan:2018pzs} 
\begin{equation}\label{517}
G^{dS}(Z(x,x')) = \Big(\frac{H^2}{16\pi^2}\Big)\Big(\frac{2}{\delta}+ \frac{4}{y} -4 - 2ln(y) + 4ln2 + O(\delta)\Big) \, ,
\end{equation} 
\noindent and using the formula Eq.~\eqref{506}, we see that the response rate of transition, for $\tilde{\eta} \in (\eta_{i}, (\eta_{i} + \eta_{f})/2)$, is given by
\begin{eqnarray}\label{519}
 \frac{1}{c^2|{}_{D}\braket{\Omega|\hat{\mu}(0)|0}_{D}|^2}   \frac{dP^{dS}_{0\rightarrow \Omega}}{d\tilde{\eta}} &=&    \frac{1}{16\pi^2}\int_{-2(\tilde{\eta} - \eta_{i})}^{2(\tilde{\eta} - \eta_{i})} d(\Delta\eta) \Big(\frac{\tilde{\eta} + (\Delta \eta)/2}{\tilde{\eta} - (\Delta \eta)/2}\Big)^{\frac{i\Omega}{H}}\frac{1}{(\tilde{\eta}^2- (\Delta \eta)^2/4)} \nonumber \\ &&\Big(\frac{2}{\delta}- \frac{4(\tilde{\eta}^2- (\Delta \eta)^2/4)}{(\Delta \eta - i \epsilon)^2} -4 - 2ln\Big( -\frac{(\Delta \eta - i \epsilon)^2}{(\tilde{\eta}^2- (\Delta \eta)^2/4)}\Big) + 4ln2 + O(\delta)\Big)  \, .
\end{eqnarray}
\noindent Using the expression Eq.~\eqref{507}, we obtain a similar formula for the response rate when $\tilde{\eta} \in ((\eta_{i} + \eta_{f})/2, \eta_{f})$. We notice that, for the above integral, the integrand has poles at $(\Delta \eta) = \pm 2\tilde{\eta}, i\epsilon$ and the interval over which the integral is performed does not include the $\pm \tilde{\eta}$ poles. The above integral, for any term in the integrand, can be easily seen to be finite in value by enclosing the contour in the lower half plane (as shown in Fig.~\ref{cont}) and noting that the value of the above integral is just equal to the integral of the above integrand along the part of the contour lying in the lower half plane, with no contribution coming from the pole lying in the upper half plane. The value of the integral of the above integrand along the part of the contour lying in the lower half plane is finite as the integral is a proper integral (By proper integral, we mean that the integral is over a finite interval and the integrand never diverges for any point along the interval and hence, these integrals are always finite). Since the integral along the curved part is a proper integral, the integral of all terms appearing in the above expression are finite. With these facts in hand, we see that the dominant contribution to the above expression in the $\delta \to 0$ limit comes from the  $1/\delta$ term. Hence, for $\delta$ very close to zero, the most dominant term to the response rate is given by  
\begin{eqnarray}\label{520}
 \frac{1}{c^2|\braket{\Omega|\hat{\mu}(0)|0}|^2}   \frac{dP^{dS}_{0\rightarrow \Omega}}{d\tilde{\eta}} &=&   \frac{1}{\delta}\Bigg(\frac{1}{8\pi^2}\int_{-2(\tilde{\eta} - \eta_{i})}^{2(\tilde{\eta} - \eta_{i})} d(\Delta\eta) \Big(\frac{\tilde{\eta} + (\Delta \eta)/2}{\tilde{\eta} - (\Delta \eta)/2}\Big)^{\frac{i\Omega}{H}}\frac{1}{(\tilde{\eta}^2- (\Delta \eta)^2/4)}\Bigg) + O(\delta^{0}).
\end{eqnarray}
From the above expression, we see that the response rate of transition for a UdW detector coupled with nearly massless scalar fields in de Sitter spacetime manifests the same infrared divergence as the Wightman function (see Eq.~\eqref{517}) for these fields. Thus, as the mass of the field decreases further, the transitions within a UdW detector occur at more faster rate. Before we turn to the case of massless scalar fields in radiation dominated spacetime, let us mention some previous works which have take up similar studies. \\ 
A number of previous works have analysed the dynamics of quantum scalar fields in de Sitter spacetime by coupling them with UdW detectors. For example, \cite{Garbrecht:2004du} has calculated the infinite time response rate (with respect to cosmic time) of UdW detectors (moving along comoving trajectories) coupled with scalar fields in de Sitter spacetime. \cite{Fukuma:2013uxa} studies UdW detectors in open quantum systems framework and place the scalar field in states other than the Bunch Davies like vacua. \cite{Tian:2013lna,Tian:2014jha} consider UdW detectors along static and free falling trajectories coupled with conformally coupled scalar fields in de Sitter spacetime and analyse them again in the open quantum systems framework. \cite{Ali:2020gij} calculates the response rate of UdW detectors (moving along comoving trajectories) which are quadratically coupled with complex scalar fields (placed in alpha vacua) in de Sitter spacetime. We extend the analysis into other FRW spacetimes which mimic different epochs of expansion of universe.


\figcontour
\subsection{Massless Scalar Fields in Radiation Dominated Spacetime}
\noindent Let us consider the case of a massless scalar field in radiation dominated universe which is understood to succeed the inflationary near de Sitter phase. For this case the scale factor, $a(\eta) = (H\eta)$ i.e., $q = -1$. Thus, the mass of the corresponding scalar field in de Sitter spacetime is $(m^2/H^2) = 2$ and the Wightman function for this case is given by
\begin{equation}\label{513}
G^{rad}(x(\eta_{1}),x(\eta_{2})) = (H^2\eta_{1}\eta_{2})^{-2}\frac{H^2}{4\pi^2 y(x(\eta_{1}),x(\eta_{2}))} = -(H^2\eta_{1}\eta_{2})^{-1}\frac{1}{4\pi^2(\eta_{1}-\eta_{2}-i\epsilon)^2} \, .
\end{equation}
This case is not just conformally related to a massive scalar field in de Sitter spacetime by the above discussed equivalence but it is also conformally related to a massless scalar field in flat spacetime (see Chapter 3 of \cite{Birrell:1982ix}). In fact, we see that the above Wightman function of massless scalar field in radiation dominated spacetime is conformally related to the Wightman function of a massless scalar field in flat spacetime. For radiation dominated case, the comoving time and the conformal time coordinates are related by the relation $(2Ht)^{\frac{1}{2}} = H\eta$. 
Using the above expression for the Wightman function and the relation between comoving and conformal coordinates, we find that the transition probability for this case is given by 
\begin{eqnarray}\label{514}
    P^{rad}_{0\rightarrow \Omega} &=& -\frac{c^2|{}_{D}\braket{\Omega|\hat{\mu}(0)|0}_{D}|^2}{4\pi^2}\int^{\eta_{f}}_{\eta_{i}}\int^{\eta_{f}}_{\eta_{i}}d\eta_{1} d\eta_{2} e^{-\frac{i\Omega H}{2}(\eta^2_{1}-\eta^2_{2})}\frac{1}{(\eta_{1}-\eta_{2} - i\epsilon)^2} \, .
\end{eqnarray}
Now we use the formula, Eq.~\eqref{506}, for rate of transition with respect to $\tilde{\eta}$ and obtain that, for $\tilde{\eta} \in \big(\eta_{i},(\eta_{i} + \eta_{f})/2\big)$, it is given by 
\begin{eqnarray}\label{515}
 \frac{1}{c^2|{}_{D}\braket{\Omega|\hat{\mu}(0)|0}_{D}|^2}   \frac{dP^{rad}_{0\rightarrow \Omega}}{d\tilde{\eta}} &=&    -\frac{1}{4\pi^2}\int_{-2(\tilde{\eta} - \eta_{i})}^{2(\tilde{\eta} - \eta_{i})} d(\Delta\eta) e^{-i\Omega H\tilde{\eta}(\Delta \eta)}\frac{1}{(\Delta \eta -i\epsilon)^2} \, .
\end{eqnarray}
One obtains a similar formula for $\tilde{\eta} \in ((\eta_{i} + \eta_{f})/2,\eta_{f})$ by using the expression Eq.~\eqref{507}. It is clear, from the form of the above expression, that the response rate for massless scalar fields in radiation dominated spacetime is similar to that for flat spacetime except for the fact that the $\Omega$ dependence in flat spacetime is replaced by $\Omega H\tilde{\eta}$ in radiation case. The above integral is over a finite interval with a pole at $i\epsilon $ lying in the upper half $\Delta \eta$ complex plane. Like in the previous subsection, we close the contour from the lower half $\Delta \eta$ complex plane and hence the contour does not contain any pole inside it. Thus, the value of the above integral along the specified interval on the real line is equal to the integral along the curved part of the contour which lies in the lower half plane. Thus, we see that the response rate for a UdW detector coupled with a massless scalar field in radiation dominated spacetime is finite and there is no substantial enhancement. Now we analyse the case of massless scalar fields in nearly matter dominated spacetimes. 
\subsection{Massless Scalar Fields in Nearly Matter Dominated Spacetimes}
\begin{figure}[h]
        \centering
        \subfloat[For $q \in (-2,0)$\newline \newline (Taking $\eta_{i} = 1, \eta_{f} = 4, \tilde{\eta} = 2$, $\Omega = 1$ and $H=1$)]{\includegraphics[scale=0.50]{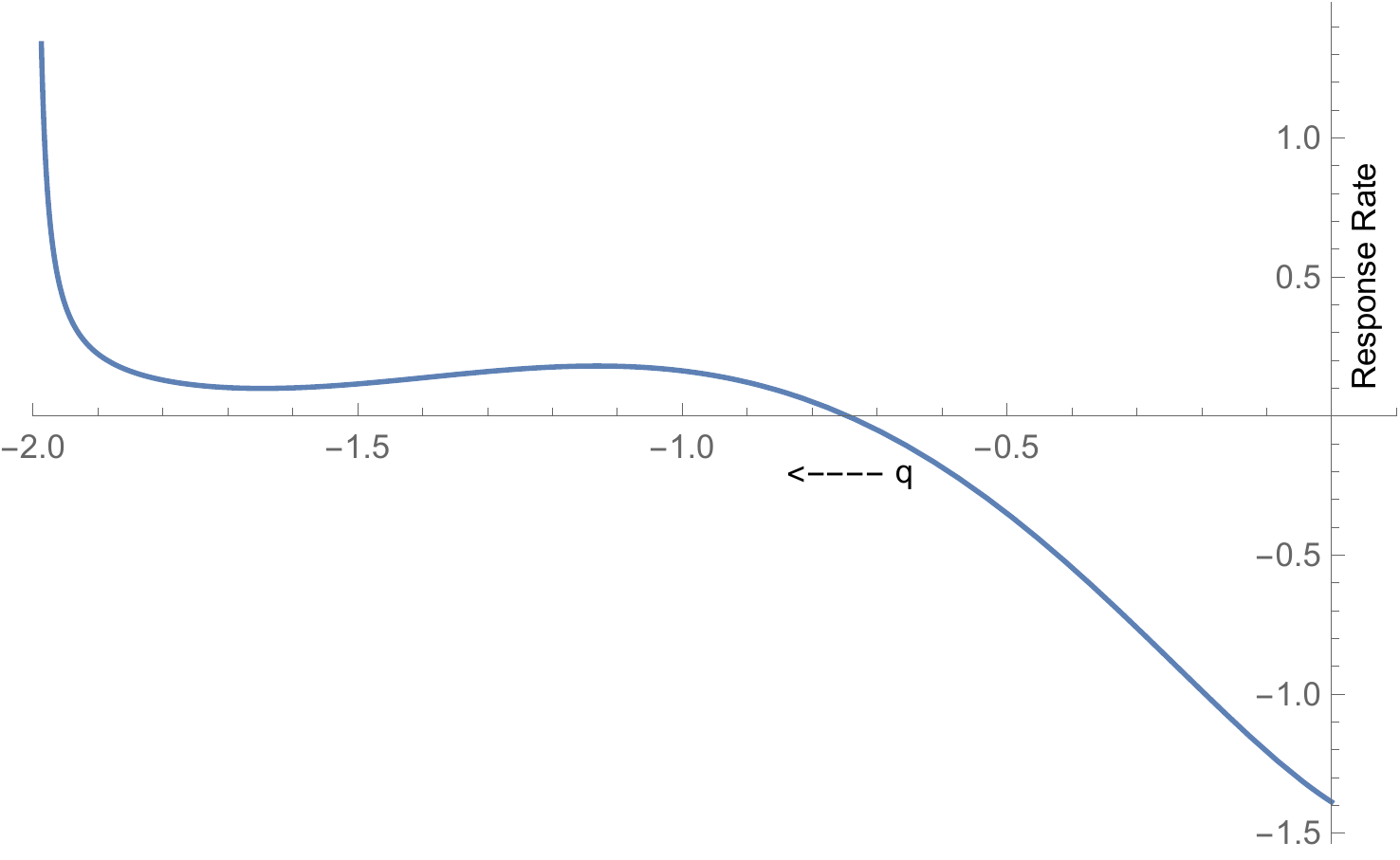}\label{rr_vs_q_1}}%
        \qquad
        \subfloat[For $q \in (0,1)$ \newline \newline (Taking $\eta_{i} = -3, \eta_{f} = 0, \tilde{\eta} = -2$, $\Omega = 1$ and $H=1$) ]{\includegraphics[scale=0.52]{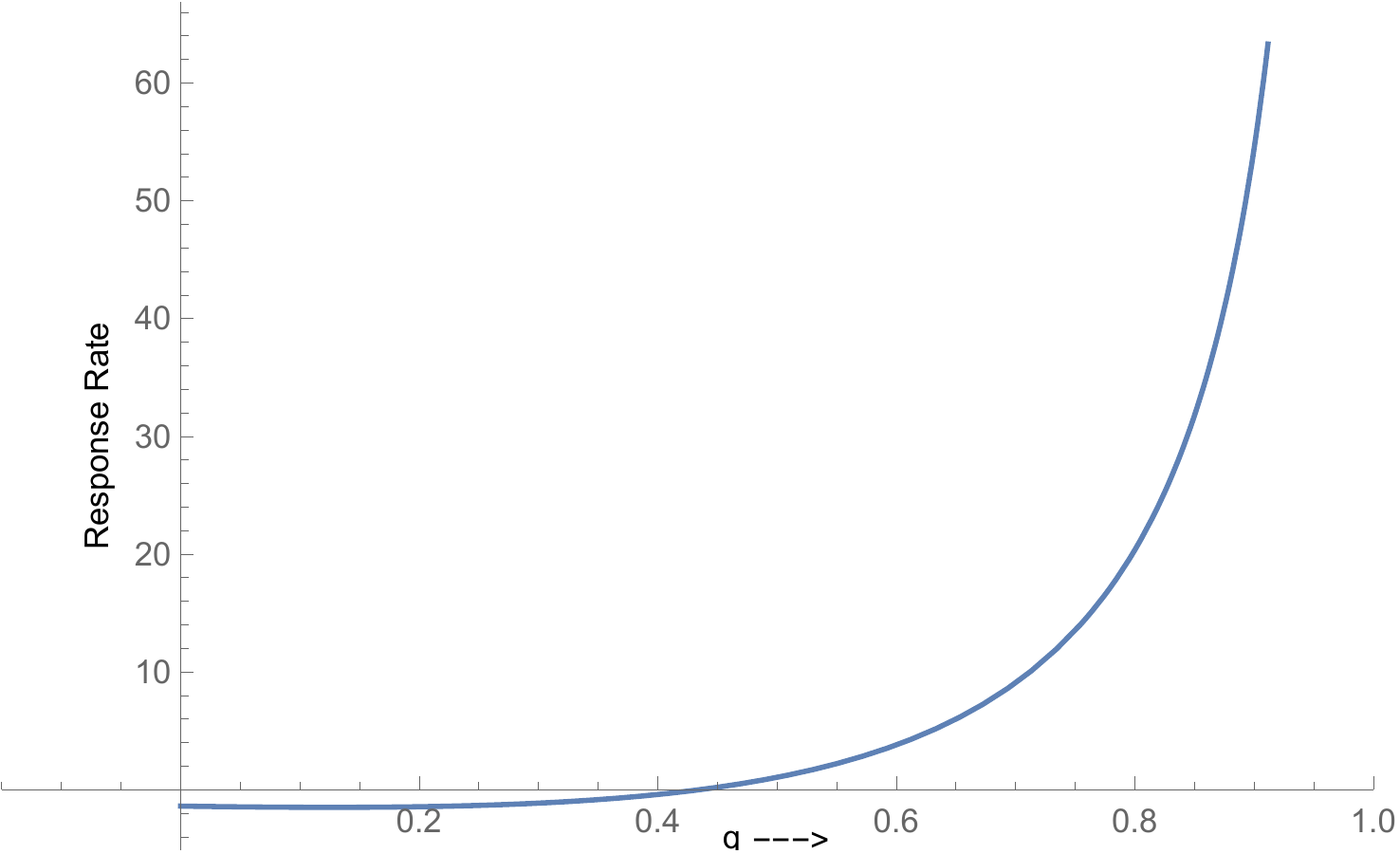}\label{rr_vs_q_2}}%
        
        \caption{Variation of the response rate for a conventionally coupled UdW detector as a function of $q$. (a) shows that the response rate diverges as $q$ approaches $-2$ i.e., matter dominated spacetime and similarly (b) shows that the response rate diverges as $q$ approaches $1$ i.e., de Sitter spacetime.    }\label{UdW_var}
    \end{figure}

\noindent In this subsection, we look at the behaviour of the response rate for UdW detectors which couple with massless scalar fields in nearly matter dominated spacetimes. Since for nearly matter dominated spacetimes i.e., for $q = -2 + \delta$ where $\delta << 1$, the mass of the corresponding scalar field in de Sitter spacetime is given by 
\begin{equation}\label{521}
\frac{m^2}{H^2} = (1-q)(2 + q) = (3 - \delta)\delta \approx 3 \delta \, ,
\end{equation} 
and approaches to zero in the $\delta$ going to zero limit which maps massless in matter dominated case to massless in de Sitter, we expect that the response rate for nearly matter dominated spacetimes inherit the infrared divergence behaviour from their counterpart of nearly massless fields in de Sitter spacetime. To investigate this case, let us write down the Wightman function of massless scalar fields in nearly matter dominated spacetimes which, using Eqs.~\eqref{510}-\eqref{512} and Eq.~\eqref{521}, is given by
\begin{eqnarray}\label{522}
G^{matter}(x(\eta_{1}),x(\eta_{2})) &=& (H^2\eta_{1}\eta_{2})^{-3 + \delta}\Big(\frac{H^2}{16\pi^2}\Big)\Big(\frac{2}{\delta}+ \frac{4}{y} -4 - 2ln(y) + 4ln2 + O(\delta)\Big) \, . 
\end{eqnarray}
Using the relation between comoving and conformal coordinates i.e.,  $\big((3-\delta)Ht\big)^{\frac{1}{3-\delta}} = (H\eta)$, in nearly matter dominated spacetimes and the formula Eq.~\eqref{506}, we see that the rate of transition probability, for $\tilde{\eta} \in (\eta_{i}, (\eta_{i} + \eta_{f})/2)$, is given by  
\begin{eqnarray}\label{523}
 \frac{1}{c^2|{}_{D}\braket{\Omega|\hat{\mu}(0)|0}_{D}|^2}   \frac{dP^{matter}_{0\rightarrow \Omega}}{d\tilde{\eta}} &=&    \frac{1}{16\pi^2}\int_{-2(\tilde{\eta} - \eta_{i})}^{2(\tilde{\eta} - \eta_{i})} d(\Delta\eta) e^{-\frac{i\Omega H^2(3\tilde{\eta}^2 + (\Delta \eta)^2/(4))(\Delta\eta)}{3}}\frac{1}{(\tilde{\eta}^2- (\Delta \eta)^2/4)} \nonumber \\ & & \Bigg(\frac{2}{\delta}- \frac{4(\tilde{\eta}^2- (\Delta \eta)^2/4)}{(\Delta \eta - i \epsilon)^2} - 4 - 2ln\Big( -\frac{(\Delta \eta - i \epsilon)^2}{(\tilde{\eta}^2- (\Delta \eta)^2/4)}\Big) + 4ln2 \nonumber \\ & &  -\frac{2i\Omega H^2 (\tilde{\eta}+ (\Delta \eta)/2)^3}{3}\Big(\frac{1}{3}-log(H(\tilde{\eta}+(\Delta \eta)/2))\Big) \nonumber \\ &&  + \frac{2i\Omega H^2 (\tilde{\eta}- (\Delta \eta)/2)^3}{3}\Big(\frac{1}{3}-log(H(\tilde{\eta}-(\Delta \eta)/2))\Big) + O(\delta)\Bigg)   \, .
\end{eqnarray}
Different terms in the integrand have same type of pole structure as in the de Sitter case and hence we can argue the finiteness of each term in the above expression just as we have done for the de Sitter case. We see that the leading order term in the $\delta \to 0$ limit is given by 
\begin{eqnarray}\label{524}
 \frac{1}{c^2|{}_{D}\braket{\Omega|\hat{\mu}(0)|0}_{D}|^2}   \frac{dP^{matter}_{0\rightarrow \Omega}}{d\tilde{\eta}} &=&    \frac{1}{\delta}\Bigg(\frac{1}{8\pi^2}\int_{-2(\tilde{\eta} - \eta_{i})}^{2(\tilde{\eta} - \eta_{i})} d(\Delta\eta) e^{-\frac{i\Omega H^2(3\tilde{\eta}^2 + (\Delta \eta)^2/(4))(\Delta\eta)}{3}}\frac{1}{(\tilde{\eta}^2- (\Delta \eta)^2/4)} \Bigg) + O(\delta^{0})\, . \nonumber \\ &&  
\end{eqnarray}
\noindent From the above expression, we see that the rate of transitions for a UdW detector which couples with massless scalar fields in nearly matter dominated spacetimes is dominated by the $1/\delta$ term in the $\delta \to 0$ limit. Thus, for $\delta$ being close to zero, we expect that the transitions within the internal quantum states of a UdW detector would take place at a rapid rate. As mentioned before, this behaviour of a conventionally coupled UdW detector in this case finds its origin in the infrared divergence of the corresponding nearly massless scalar fields in de Sitter spacetime. In fact, if we look at the Eqs. (\ref{510}) and (\ref{511}) for the Wightman function of a massless scalar field in FRW spacetimes with $q \in [-2,1]$, we expect that a UdW detector should experience very fast transitions between its quantum states as one approaches both $q = -2$ and $q =1$ cases which is what we have seen in this and the previous subsection.  In order to demonstrate the behaviour of UdW detectors in the considered spacetimes with $q\in (-2,1)$, we plot the response rate expression, Eq.~\eqref{506}, for the Wightman function Eq.~\eqref{511} as a function of $q$ for specific values of $\eta_{i}$, $\eta_{f}$, $\tilde{\eta}$, $\Omega$ and $H$. For example, Fig.~\ref{rr_vs_q_1} shows the variation of response rate as $q$ is varied between $(-2,0)$ and Fig.~\ref{rr_vs_q_2} shows the variation of the response rate as $q$ is varied between $(0,1)$. As expected, we observe from the figures that the response rate remains finite except for the cases when $q$ approaches the values $-2$ and $1$ 

\noindent We have seen above that the infrared divergence of massless scalar fields in de Sitter spacetime manifests itself at the level of response rate of UdW detectors. In fact, all the quantities that involve the Wightman function can be, in general, expected to contain these infrared divergences. These divergences have been a subject of much discussion \cite{Vilenkin:1982wt, Antoniadis:1985pj, Allen:1987tz,Polarski:1991ek,Dolgov:1994cq,Miao:2010vs}. The fact that one can not have a well defined vacuum state for a massless scalar field in de Sitter spacetime which enjoys the full de Sitter symmetry is well known and in order to find physically sensible results for this case a number of suggestions have been made in the literature \cite{Marolf:2010zp,Dolgov:1994cq}. For example, to obtain results free of infrared divergences, sometimes less symmetric states are considered as vacua for massless scalar fields in de Sitter spacetime \cite{Vilenkin:1982wt, Allen:1985ux, Allen:1987tz,Kirsten:1993ug}. Another line of thought regarding the possible resolution of these infrared divergences is to argue that only those operators be considered physical which are infrared finite. For example, \cite{Page:2012fn} suggests that one should consider only shift invariant operators, like the differences of the field operators and derivatives of the field operators etc., as true physical observables do not suffer from infrared divergences. Introduction of a small mass not only regularizes the correlator into one containing a finite (but large) constant term, it further removes this constant term from any physical operators. For instance, \cite{Ford:1984hs}  points out that the stress energy expectation does not suffer from infrared divergence because it contains derivatives which remove the infrared divergences (or the large constant terms in small mass limit). On the basis of these previous considerations, we expect that if we consider more `physical' derivatively coupled UdW detectors, then the corresponding response rates should be free of infrared divergences. In fact, there have been previous works which have considered derivatively coupled UdW detectors in order to deal with  the infrared divergences that appear in different contexts \cite{Tjoa:2020eqh, Tjoa:2022oxv, Juarez-Aubry:2014jba}. Motivated by this, in the next section, we consider the case of derivatively coupled UdW detectors which couple to massless scalar fields in FRW spacetimes. We will see that even though the infrared divergence for de Sitter spacetime does not contribute to the response rate of these detectors, the infrared divergence present in a matter dominated spacetime still contributes to the response rate of these detectors.  Even for a nearly matter dominated era, the correlation function contains a large term multiplied to the conformal factor through which it survives under derivative actions. Potential presence of such divergent terms in correlators in matter dominated era leads to observable signatures such as a large response rate of the derivatively coupled UdW detectors, which we see next.

\section{Derivatively coupled UdW detectors}\label{dUdW}
\noindent In this section, we analyse derivatively coupled UdW detectors. A derivatively coupled UdW detector is just like the conventional UdW detector except for the form of the interaction with the quantum field which, in this case, is given by the following Hamiltonian \cite{Juarez-Aubry:2014jba}
\begin{equation}\label{525}
H_{int} = c\hat{\mu}(\tau)\chi(\tau)\dot{x}^{\sigma}\nabla_{\sigma}\hat{\phi}(x(\tau)) = c\hat{\mu}(\tau)\chi(\tau)\frac{d}{d\tau}\hat{\phi}(x(\tau)) \, ,
\end{equation}
where all the terms have the same meaning as in the previous section and $\frac{d}{d\tau}\hat{\phi}(x(\tau))$ is the derivative of the quantum field with respect to the proper time along the classical trajectory of the detector. Here dot in $\dot{x}^{\sigma}$ also represents the derivative with respect to the proper time. If we, again, consider the case in which the detector makes a transition from some state $\ket{0}_{D}$ to another state $\ket{\Omega}_{D}$ which have energies $0$ and $\Omega$, respectively, and the field starts in some state $\ket{\psi}$ while it is allowed to go to any final state, then the transition probability for this case, in first order perturbation theory, is given by 
\begin{equation}\label{526}
 P_{0\rightarrow \Omega}=c^2|{}_{D}\braket{\Omega|\hat{\mu}(0)|0}_{D}|^2\iint d\tau_{1} d\tau_{2}e^{-i\Omega(\tau_{1}-\tau_{2})}\chi(\tau_{1})\chi(\tau_{2})\frac{d}{d\tau_{1}}\frac{d}{d\tau_{2}}G(x(\tau_{1}),x(\tau_{2})).
\end{equation}
As in the previous section we are ultimately interested in FRW spacetimes. Therefore, we can go to the conformal coordinates i.e.,  $dt = a(\eta)d\eta$, and the above expression becomes 
\begin{equation}\label{528}
    P_{0\rightarrow \Omega}=c^2|{}_{D}\braket{\Omega|\hat{\mu}(0)|0}_{D}|^2\int_{\eta_{i}}^{\eta_{f}} \int_{\eta_{i}}^{\eta_{f}} d\eta_{1} d\eta_{2}e^{-i\Omega(\tau(\eta_{1})-\tau(\eta_{2}))}\frac{d}{d\eta_{1}}\frac{d}{d\eta_{2}}G(x(\eta_{1}),x(\eta_{2})) \, .
\end{equation}

\noindent We go to the $(\tilde{\eta}, \Delta \eta)$ coordinates (defined in the previous section) to express the formulae for the rate of transition probability.  For $\tilde{\eta} \in (\eta_{i},(\eta_{i} + \eta_{f})/2)$, the response rate with respect to $\tilde{\eta}$ is given by \\
\begin{eqnarray}\label{529}
\frac{1}{c^2|{}_{D}\braket{\Omega|\hat{\mu}(0)|0}_{D}|^2}\frac{dP_{0\rightarrow \Omega}}{d\tilde{\eta}} &=& \int_{-2(\tilde{\eta} - \eta_{i})}^{2(\tilde{\eta} - \eta_{i})} d(\Delta\eta) \ e^{-i\Omega(\tau(\tilde{\eta} + (\Delta \eta)/2)-\tau(\tilde{\eta} - (\Delta \eta)/2))} \nonumber \\ &&  \Big[\Big(\frac{d}{d\eta_{1}}\frac{d}{d\eta_{2}}G\Big)\Big(x\big(\tilde{\eta} + (\Delta \eta)/2\big),x\big(\tilde{\eta} - (\Delta \eta)/2\big)\Big)\Big] \, ,
\end{eqnarray}
whereas for $\tilde{\eta} \in ((\eta_{i} + \eta_{f})/2, \eta_{f})$, the response rate is given by \\
\begin{eqnarray}\label{530}
\frac{1}{c^2|{}_{D}\braket{\Omega|\hat{\mu}(0)|0}_{D}|^2}\frac{dP_{0\rightarrow \Omega}}{d\tilde{\eta}} &=& \int_{-2( \eta_{f}-\tilde{\eta} )}^{2(\eta_{f} - \tilde{\eta} )} d(\Delta\eta)e^{-i\Omega(\tau(\tilde{\eta} + (\Delta \eta)/2)-\tau(\tilde{\eta} - (\Delta \eta)/2))} \nonumber \\ &&  \Big[\Big(\frac{d}{d\eta_{1}}\frac{d}{d\eta_{2}}G\Big)\Big(x\big(\tilde{\eta} + (\Delta \eta)/2\big),x\big(\tilde{\eta} - (\Delta \eta)/2\big)\Big)\Big] \, .
\end{eqnarray}
Here $\Big[\Big(\frac{d}{d\eta_{1}}\frac{d}{d\eta_{2}}G\Big)\Big(x\big(\tilde{\eta} + (\Delta \eta)/2\big),x\big(\tilde{\eta} - (\Delta \eta)/2\big)\Big)\Big]$ denotes that we first calculate the double derivative with respect to $\eta_{1}$ and $\eta_{2}$ and then express the resultant expression in $(\tilde{\eta}, \Delta \eta)$ coordinates.\\ \\

\noindent Let us now analyse the behaviour of the response rate for these derivatively coupled UdW detectors for the same cases which are considered in the previous section. For this purpose, we notice (refer Appendix \ref{dds}) that
 \begin{eqnarray}\label{531}
\frac{d}{d\eta_{1}}\frac{d}{d\eta_{2}}G^{FRW}(x(\eta_{1}),x(\eta_{2})) &=& (H^2\eta_{1}\eta_{2})^{q-1}\bigg[(q-1)^2\frac{G^{dS}}{\eta_{1}\eta_{2}} + (q-1)\frac{dG^{dS}}{dy}\Big(\frac{(\eta_{1}-\eta_{2}-i\epsilon)(-2i\epsilon)}{\eta_{1}^{2}\eta_{2}^{2}}\Big) \nonumber\\ & & + \frac{d^2G^{dS}}{dy^2}\frac{y((\eta_{1}+\eta_{2})^2 + \epsilon^2)}{\eta_{1}^{2}\eta_{2}^{2}} + \frac{dG^{dS}}{dy}\frac{(\eta_{1}^{2}+\eta_{2}^{2} + \epsilon^2)}{\eta_{1}^{2} \eta_{2}^{2}}\bigg]  \, .
\end{eqnarray}
Making use of the above expression, we study the behaviour of the response rate for derivatively  coupled UdW detectors and compare them with the analogous behaviour of the response rate for conventional UdW detectors investigated in the previous section. \\ \\
\noindent Before we come to finite time response rate of these derivatively coupled UdW detectors, we discuss the $\Omega$ and $H$ dependences of the infinite time response rate of these detectors. It is argued in Appendix \ref{infres} that the infinite time response rate for derivatively coupled UdW detectors in FRW spacetimes has the following $\Omega$ and $H$ dependences
\begin{equation}
\frac{1}{c^2|{}_{D}\braket{\Omega|\hat{\mu}(0)|0}_{D}|^2}\frac{dP_{0\rightarrow \Omega}}{d\tilde{\eta}} \propto \Omega^2(\Omega H^{-q})^{\frac{1}{1-q}} \, .
\end{equation} 
Thus, we see that the $\Omega$ dependence of the response rate has an extra $\Omega^2$ factor compared to the conventional UdW detector coupling. But as far as the $H$ dependence goes, it is the same as in the case of conventionally coupled UdW detectors. For $q \in (-2,0)$, the response rate increases with increasing value of $H$ whereas for $q \in (0,1)$, the response rate decreases with increasing value of $H$. Since the Ricci scalar, $R \propto H^{2q}$, this implies that for the considered spacetimes, the Ricci scalar and the response rate behave opposite to each other as a function of $H$. Like in the previous section, the above expression tells that the rates are maximally enhanced by $H$ for de Sitter and matter dominated cases depending upon whether $H<1$ or $>1$. 
In fact, in the derivatively coupling case, the proportionality constant contains the infrared divergent term only for matter dominated spacetime but not for de Sitter spacetime. Hence, unlike in the previous section, it is only the matter dominated case which shows the divergently rapid rate but not the de Sitter case. To investigate it further, let us now turn to the finite time response rates for these derivatively coupled detectors.\\
\subsection{Nearly Massless Scalar Fields in de Sitter Spacetime}
\noindent Let us now analyse the response rate of UdW detectors which are derivatively coupled with nearly massless scalar fields in de Sitter spacetime. 
\noindent Since it is the double derivative of the Wightman function which enters the expression for the response rate, Eq.~\eqref{528}, and the fact that the $1/\delta$ term in the Wightman function, Eq.~\eqref{517}, does not have any spacetime dependence, this implies that the $1/\delta$ term (which is responsible for the divergent response rate for conventional UdW detectors in the vanishing mass limit of this setting) does not contribute to the expression of the response rate for derivatively coupled UdW detectors. To obtain the response rate, we substitute Eq.~\eqref{531} in Eq.~\eqref{529} with $q=1$ and $G^{dS}$ given by Eq.~\eqref{517}. Doing that we find that the rate of transition probability, for $\tilde{\eta} \in (\eta_{i}, (\eta_{i} + \eta_{f})/2)$, is given by
\begin{eqnarray}\label{534}
 \frac{1}{c^2|{}_{D}\braket{\Omega|\hat{\mu}(0)|0}_{D}|^2}   \frac{dP^{dS}_{0\rightarrow \Omega}}{d\tilde{\eta}} &=&   \frac{H^2}{4\pi^2}\int_{-2(\tilde{\eta} - \eta_{i})}^{2(\tilde{\eta} - \eta_{i})} d(\Delta\eta) \Big(\frac{\tilde{\eta} + (\Delta \eta)/2}{\tilde{\eta} - (\Delta \eta)/2}\Big)^{\frac{i\Omega}{H}}\frac{1}{(\Delta \eta-i\epsilon)^4}\Big(6 (\tilde{\eta}^2- (\Delta \eta)^2/4)\nonumber \\ && \hspace{2cm} + 2\epsilon^2 + 2i\epsilon(\Delta \eta)\Big)   + O(\delta)  \, .
\end{eqnarray}
\noindent Hence, the response rate for this case does not inherit the infrared divergence of the massless scalar fields in de Sitter spacetime and the derivatively coupled UdW detectors do not experience divergently fast transitions among their internal quantum levels as the mass of the field approaches zero. As in the previous section, we demonstrate the behaviour of the response rate for this case by numerically plotting its expression, Eq.~\eqref{534}, for some specific values of the parameters entering in the expression. To that end, for the considered setting, Fig.~\ref{UdW_dS} shows the variation of the response rate as a function of the energy gap, $\Omega$, for a UdW detector which is derivatively coupled with a field for which $\delta = 0.001$. In the plot, negative values of $\Omega$ correspond to the de-excitation of the detector from higher energy levels to the lower energy levels while the positive values of $\Omega$ correspond to the opposite case.  The absence of the signature of infrared divergence of massless scalar fields in de Sitter spacetime in the above response rate is expected for `physical' operators like derivative operators \cite{Page:2012fn}. But, as we will see below, the infrared divergence of massless scalar fields in certain FRW spacetimes may still show up even with these `physical' derivative couplings.  
%
%
%
\subsection{Massless Scalar Fields in Radiation Dominated Spacetime}
\begin{figure}[h]
        \centering
        \subfloat[ For a nearly massless field in de Sitter spacetime\newline \newline (Taking $\eta_{i} = -3, \eta_{f} = 0, \tilde{\eta} = -2$, $H=1$ and $\delta =0.001 $) ]{\includegraphics[scale=0.52]{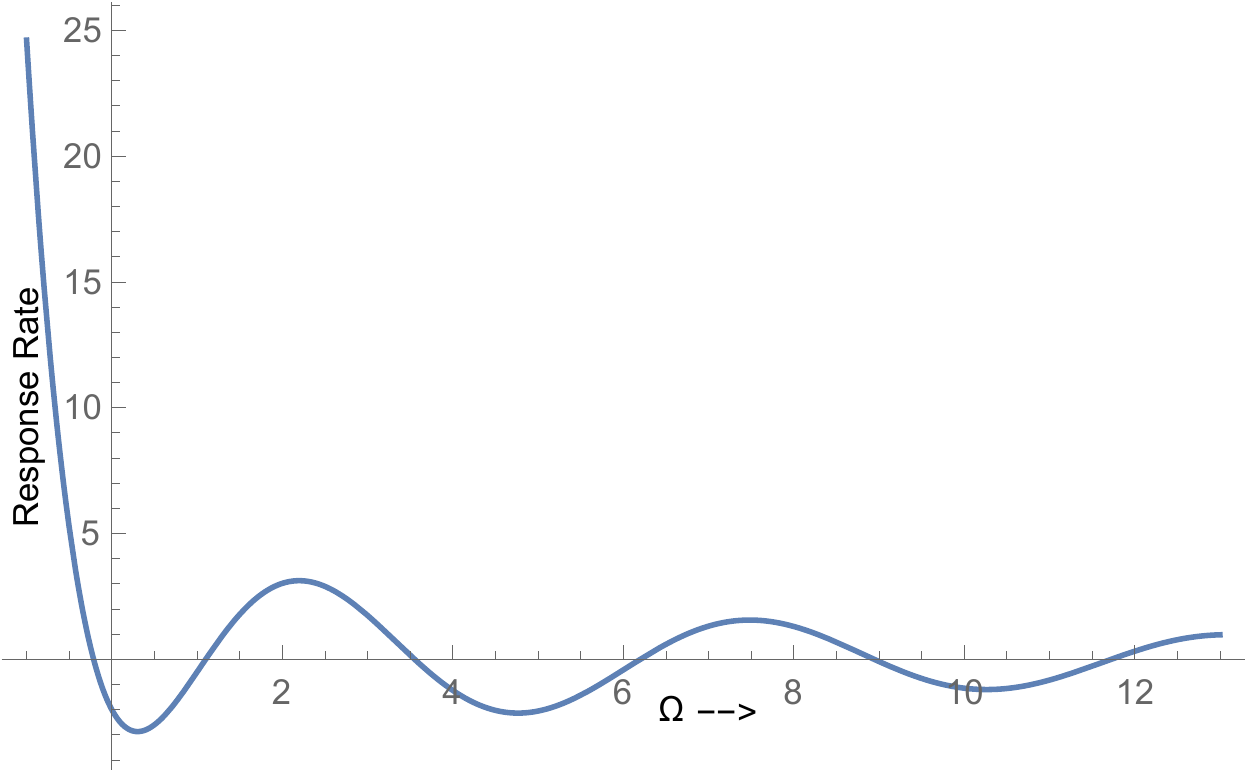}\label{UdW_dS}}%
         \qquad
        \subfloat[For a massless field in radiation dominated spacetime \newline \newline (Taking $\eta_{i} = 1, \eta_{f} = 4, \tilde{\eta} = 2$ and $H=1$)]{\includegraphics[scale=0.50]{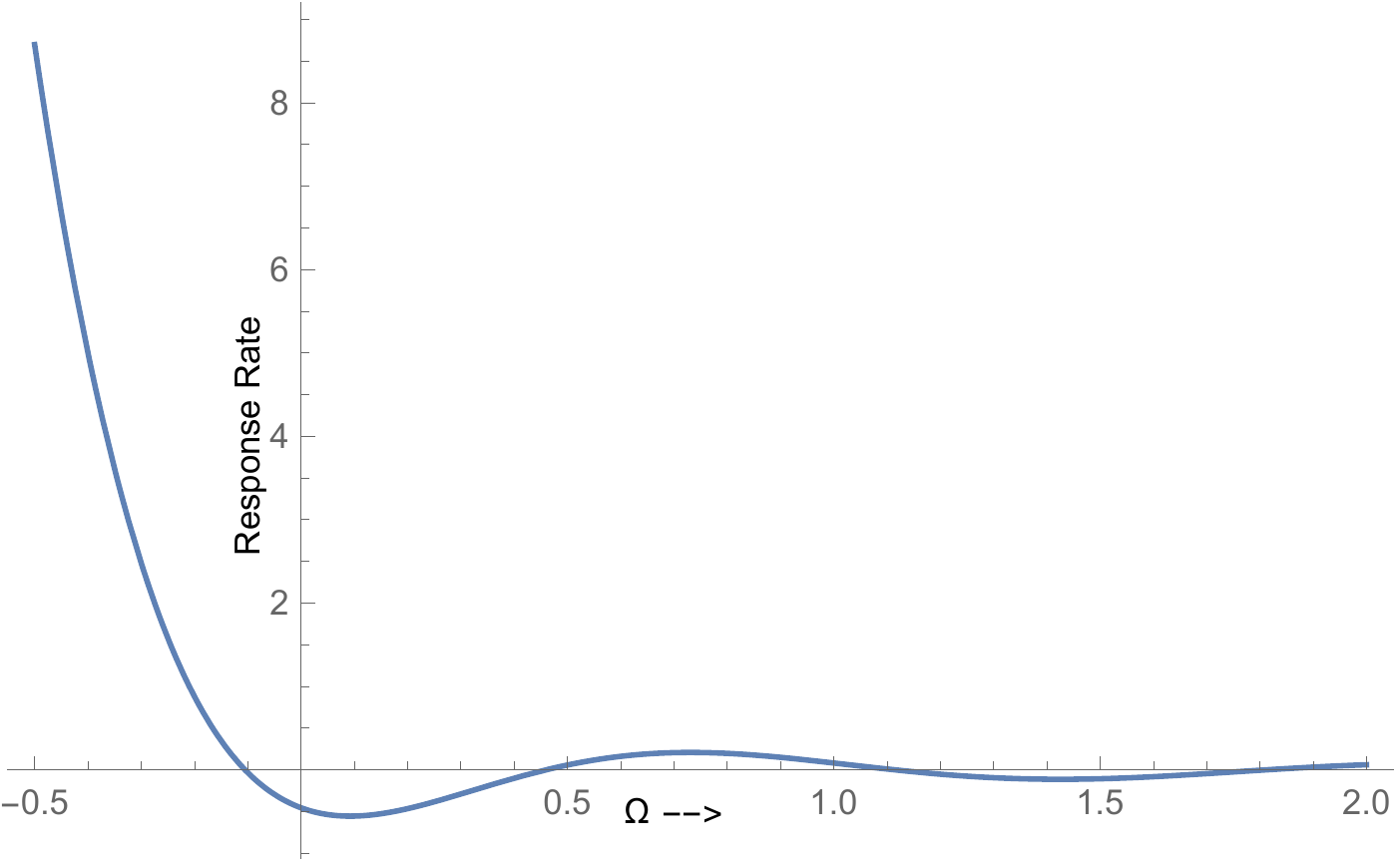}\label{UdW_Rad}}%
        \qquad
                \subfloat[For a massless field in a nearly matter dominated spacetime \newline \newline (Taking $\eta_{i} = 1, \eta_{f} = 4, \tilde{\eta} = 2$, $H=1$ and $\delta =0.001 $) ]{\includegraphics[scale=0.52]{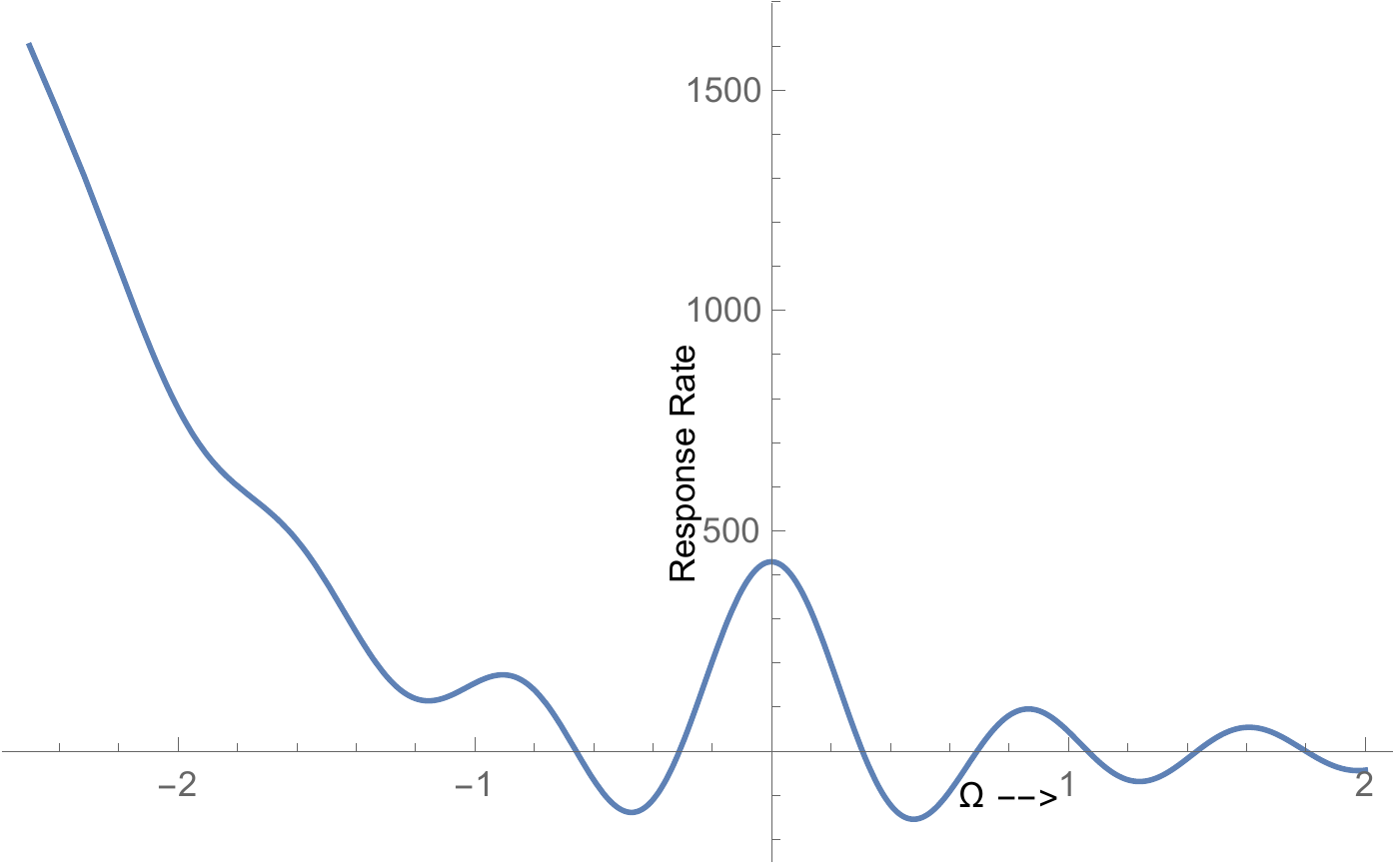}\label{UdW_Mat}}%
        \caption{Variation of the response rate for a derivatively coupled UdW detector as a function of $\Omega$.   }%
        \label{dUdW_va}%
    \end{figure}

\noindent In this subsection, we consider derivative coupling of UdW detectors with massless scalar fields in radiation dominated spacetime i.e., $q = -1$ case. Therefore, using Eq.~\eqref{531}, we obtain 
\begin{eqnarray}\label{532}
\frac{d}{d\eta_{1}}\frac{d}{d\eta_{2}}G^{rad}(x(\eta),x(\eta')) &=& (H^2\eta_{1}\eta_{2})^{-2}\bigg[-4\frac{H^2}{4\pi^2 (\eta_{1}-\eta_{2}-i\epsilon)^2} -\frac{4H^2i\epsilon(\eta_{1}-\eta_{2}-i\epsilon)}{4\pi^2 (\eta_{1}-\eta_{2}-i\epsilon)^4} \nonumber\\ & & + \frac{2H^2((\eta_{1}+\eta_{2})^2 + \epsilon^2)}{4\pi^2(\eta_{1}-\eta_{2}-i\epsilon)^4} - \frac{H^2(\eta_{1}^{2}+\eta_{2}^{2} + \epsilon^2)}{4\pi^2(\eta_{1}-\eta_{2}-i\epsilon)^4}\bigg]  \, .
\end{eqnarray}
The response rate with respect to $\tilde{\eta}$, for $\tilde{\eta} \in (\eta_{i}, (\eta_{i} + \eta_{f})/2)$, is given by 
\begin{eqnarray}\label{533}
  \frac{1}{c^2|{}_{D}\braket{\Omega|\hat{\mu}(0)|0}_{D}|^2}   \frac{dP^{rad}_{0\rightarrow \Omega}}{d\tilde{\eta}} &=& \int_{-2(\tilde{\eta} - \eta_{i})}^{2(\tilde{\eta} - \eta_{i})} d(\Delta\eta) \frac{e^{-i\Omega H\tilde{\eta}(\Delta \eta)}}{H^2\pi^2}\frac{1}{\Big(\tilde{\eta}^2-(\Delta \eta)^2/4\Big)^2}\bigg[-\frac{1}{ (\Delta \eta -i\epsilon)^2} -\frac{i\epsilon}{(\Delta \eta -i\epsilon)^3} \nonumber\\ & & + \frac{(4\tilde{\eta}^2 + \epsilon^2)}{2(\Delta \eta -i\epsilon)^4} - \frac{(2\tilde{\eta}^2 +(\Delta \eta)^2/2 + \epsilon^2)}{4(\Delta \eta -i\epsilon)^4}\bigg]  \, .
\end{eqnarray}
The above expression for the response rate can be argued to be finite by using essentially the same arguments as were used in the previous section. Thus, for a UdW detector which is derivatively coupled with massless scalar fields in radiation dominated spacetime, the rate of transitions within the internal quantum states of the detector are finite just like for a conventionally coupled UdW detector in the similar setting. Fig.~\ref{UdW_Rad} shows the variation of the response rate as a function of the energy gap between the levels between which the detector makes the transition. Fig.~\ref{UdW_dS} and Fig.~\ref{UdW_Rad} clearly show that the vacuum de-excitation rate for both de Sitter and radiation dominated spacetimes is more prominent compared to the excitation rate.

\subsection{Massless Scalar Fields in Nearly Matter Dominated Spacetimes}
\noindent In this subsection, we analyse the behaviour of UdW detectors which are derivatively coupled with massless scalar fields in nearly matter dominated spacetimes. 
For this case, the $1/\delta$ term has, unlike the case of nearly massless scalar fields in de Sitter spacetime, spacetime dependence and hence it does not disappear under the action of double time derivatives appearing in the expression of the response rate Eq.~\eqref{529} or Eq.~\eqref{530}.  In fact, using the expression Eq.~\eqref{522} for the Wightman function and the relation between the comoving and conformal time coordinates for this case i.e., 
\begin{equation}\label{536}
t = \frac{H^2\eta^3 e^{-\delta (ln(H\eta))}}{3-\delta} = \frac{H^2\eta^3}{3}\Big(1 +  \frac{\delta}{3}- \delta\ ln(H\eta)\Big) + O(\delta^2) \, ,
\end{equation}
in the expression Eq.~\eqref{529} for the response rate, we obtain that the response rate, for $\tilde{\eta} \in (\eta_{i}, (\eta_{i} + \eta_{f})/2)$, is given by 

\begin{eqnarray}\label{537}
\frac{1}{c^2|{}_{D}\braket{\Omega|\hat{\mu}(0)|0}_{D}|^2}   \frac{dP^{matter}_{0\rightarrow \Omega}}{d\tilde{\eta}} &=&    \frac{1}{\delta}\Bigg(\frac{9}{8 H^4 \pi^2}\int_{-2(\tilde{\eta} - \eta_{i})}^{2(\tilde{\eta} - \eta_{i})} d(\Delta\eta) e^{-\frac{i\Omega H^2(3\tilde{\eta}^2 + (\Delta \eta)^2/(4))(\Delta\eta)}{3}}\frac{1}{(\tilde{\eta}^2- (\Delta \eta)^2/4)^{4}} \Bigg) \nonumber  \\ & & + O(\delta^{0})\, . 
\end{eqnarray}
From the above expression, we see that the leading order term in the $\delta \to 0$ limit is $1/\delta$ and it leads to very large transition rates among the internal quantum states of a UdW detector. Thus, the response rate for UdW detectors which are derivatively coupled with massless fields in nearly matter dominated spacetimes manifests the infrared divergence of the corresponding nearly massless scalar fields in de Sitter spacetime. This is, in contrast, to the response rate for UdW detectors which are derivatively coupled with nearly massless fields in de Sitter spacetime where, as we saw, the infrared divergence of the massless limit case is cured by the time derivatives of the Wightman function present in the expression of the response rate Eq. (\ref{528}) and the fact that the infrared divergence, in that case, is spacetime independent. In order to demonstrate the behaviour of the response rate for the present case, we plot the response rate numerically for a set of values for the parameters which appear in its expression. Fig~\ref{UdW_Mat} shows the variation of the response rate as a function of the energy gap, $\Omega$. We observe that in this case, unlike the previous two cases where there was no symmetry between the excitation and de-excitation behaviours, the response rate shows similar behaviour both for excitations and de-excitations, at least, for small $\Omega$. This is also apparent from the expression Eq. (\ref{537}) in which the leading order term is invariant under the change of sign of $\Omega$ and thus we expect that the behaviour of the response rate should be invariant under the change of sign of $\Omega$ in the $\delta \to 0$ limit. At larger $\Omega$, the sub-dominant terms grow larger and the vacuum de-excitation again takes over breaking the $\Omega \to -\Omega$ symmetry but for small $\Omega$, the rates are symmetric to a good extent. {\it Thus the derivatively coupled UdW detector with small $\Omega$ behaves as if it is put in a large temperature thermal bath where both excitation and de-excitation rates are similar.}

\begin{figure}[h]
        \centering
        \subfloat[For $q \in (-2,0)$\newline \newline (Taking $\eta_{i} = 1, \eta_{f} = 4, \tilde{\eta} = 2$, $\Omega = 1$ and $H=1$)]{\includegraphics[scale=0.50]{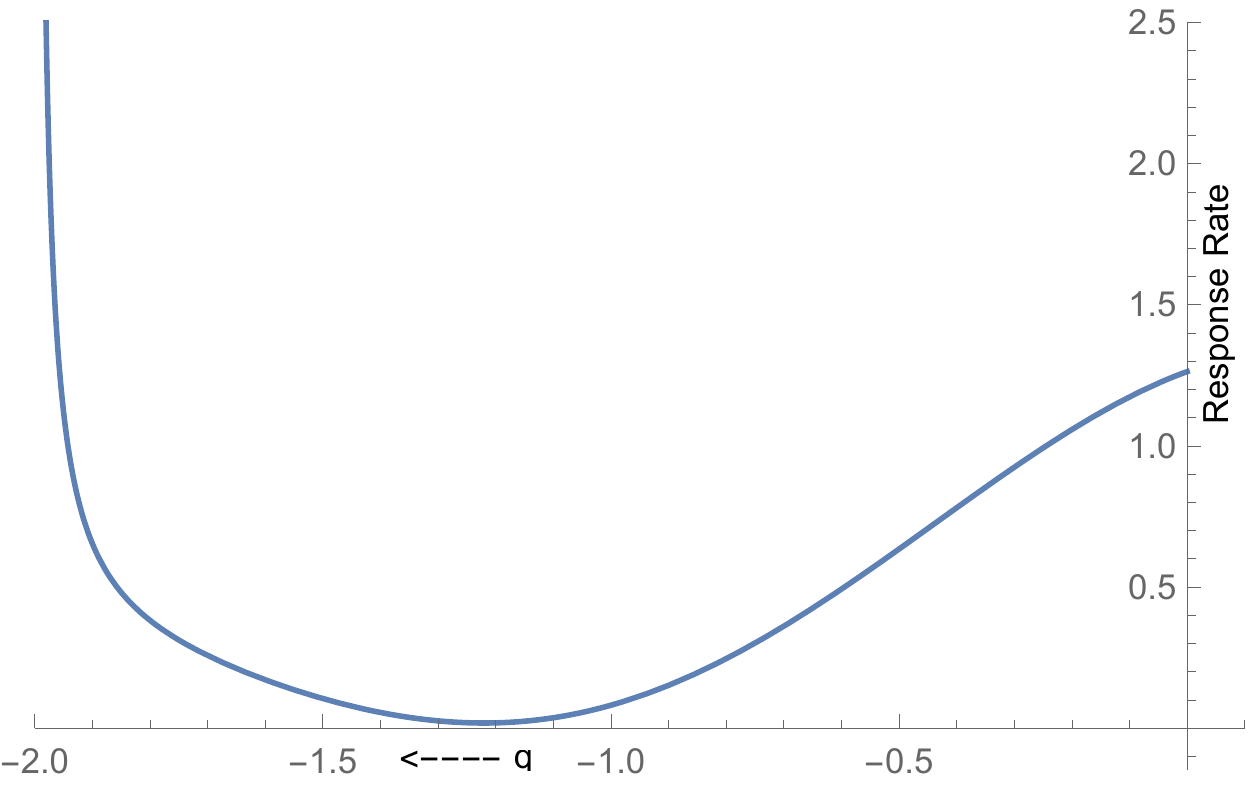}\label{dUdW_pq}}%
        \qquad
        \subfloat[For $q \in (0,1)$ \newline \newline (Taking $\eta_{i} = -3, \eta_{f} = 0, \tilde{\eta} = -2$, $\Omega = 1$ and $H=1$) ]{\includegraphics[scale=0.52]{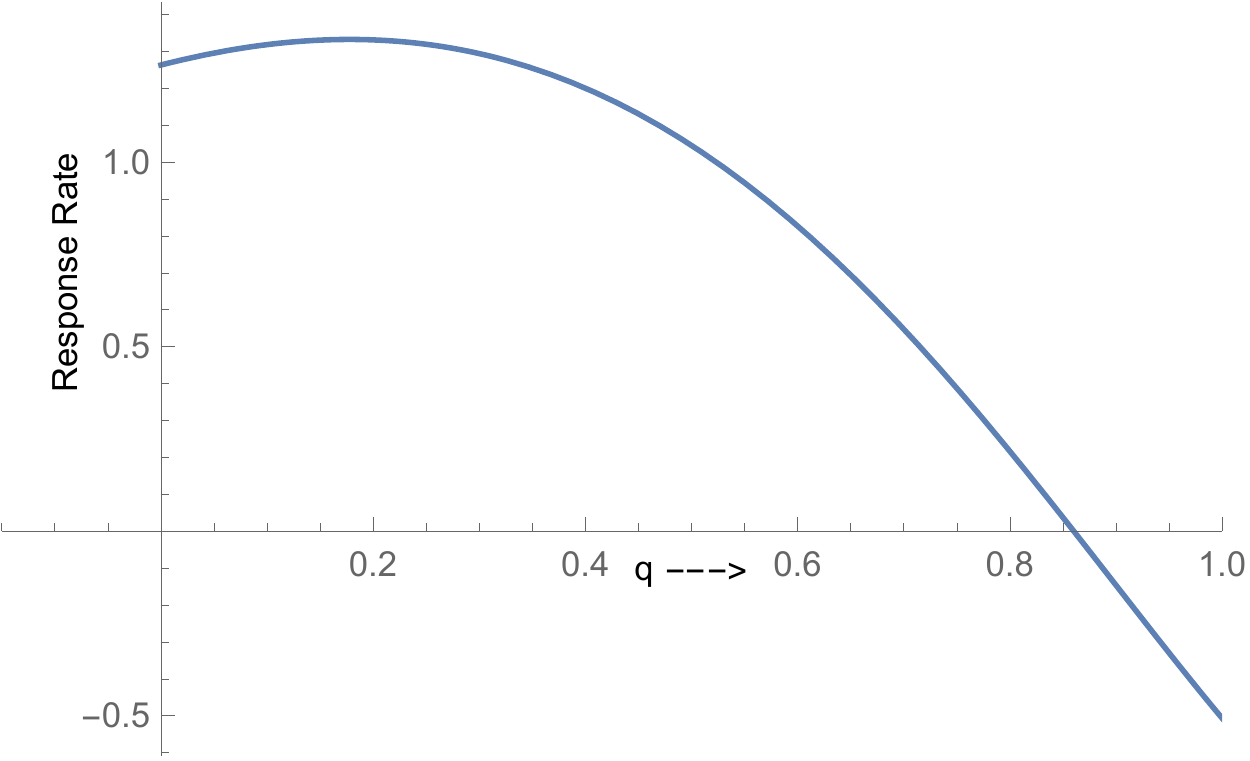}\label{dUdW_nq}}%
        
        \caption{Variation of the response rate for a derivatively coupled UdW detector as a function of q. (a) shows that the response rate diverges as $q$ approaches $-2$ i.e., derivatively coupled UdW detectors manifest infrared divergence for matter dominated spacetime. However, (b) shows that the response rate is finite for all $q \in (0,1)$ i.e., the infrared divergence of de Sitter spacetime disappears for derivatively coupled detectors.    }%
        \label{dUdW_var}%
    \end{figure}    
\noindent In Fig.~\ref{dUdW_var} we plot the variation of the response rate as a function of q for a set of values for other parameters that go into the expression of the response rate for derivatively coupled UdW detectors. We observe, as has been argued above, that the response rate shows divergent behaviour as $q$ approaches $-2$ but remains finite as $q$ approaches $1$. Like mentioned above, this is in contrast to the behaviour of the response rate for conventionally coupled UdW detectors for which the response rate diverges as $q$ approaches both $-2$ and $1$ (see Fig.~\ref{UdW_var}).  \\
This analysis of UdW detectors which are either conventionally or derivatively coupled  with massless fields in FRW spacetimes helps us gain important insights into the correlations of quantum fields in FRW spacetimes and showed that such detectors effectively capture the late time growth of quantum correlators. This study has, till now, mostly looked at the formal aspects of quantum fields in FRW spacetimes. In the next section, we discuss applicability of these results for physical systems where UdW type of coupling arises. 
\section{Implications}\label{Impli}
\subsection{Stress-energy tensor coupling with detectors}\label{str_enrg_UdW}
\noindent  
Let us look at the implications of the above study to a specific type of UdW coupling where, instead of coupling particle detectors with the quantum field, one couples them linearly with the stress energy tensor of the quantum field \cite{Padmanabhan:1987rq} i.e.,
\begin{equation}
H_{I} = c\ \chi(\tau) \hat{\mu}^{\alpha \beta}(\tau)\hat{T}_{\alpha\beta}(x(\tau))  \, ,
\end{equation}  
where $\hat{\mu}^{\alpha \beta}$ determines the transitions in the internal quantum space of the detector and $\hat{T}_{\alpha\beta}$ is the stress energy operator of the quantum field with which the detector couples by the above interaction Hamiltonian.\\
If we again consider the case in which the detector goes from some state $\ket{0}_{D}$ to another state $\ket{\Omega}_{D}$ which have energies $0$ and $\Omega$, respectively, and the field starts in some state $\ket{\psi}$ while it is allowed to go to any final state, then the transition probability for this case, in first order perturbation theory, is given by 
\begin{equation}
 P_{0\rightarrow \Omega}=c^2 {}_{D}\braket{\Omega|\hat{\mu}^{\alpha \beta}(0)|0}_{D} {}_{D}\braket{\Omega|\hat{\mu}^{\gamma \delta}(0)|0}_{D}^{*}\iint d\tau_{1} d\tau_{2}e^{-i\Omega(\tau_{1}-\tau_{2})}\chi(\tau_{1})\chi(\tau_{2})\bra{\psi}\hat{T}_{\alpha \beta}(x(\tau_1))\hat{T}_{\gamma \delta}(x(\tau_2))\ket{\psi}.
\end{equation}
Thus, we find that the transition probability (and hence, the response rate) depends on the correlations of the stress energy tensors between the points lying on the trajectory of the detector. Hence, the response rate of stress energy UdW detectors records the cumulative correlations of stress energy operators along the trajectory of the detectors. Such correlations of stress energy operators have also been previously studied for FRW spacetimes but in different contexts \cite{Dhanuka:2020yxp,Perez-Nadal:2009jcz,Hu:2008rga}.  \\ 
We now argue that the above expression leads to similar type of conclusions that we have obtained for derivatively coupled UdW detectors. For this purpose, we note that the stress energy operator for a massless scalar field in an FRW spacetime is given by 
\begin{equation}
\hat{T}_{\alpha \beta}(x) = \partial_{\alpha}\hat{\phi}(x)\partial_{\beta}\hat{\phi}(x) - \frac{1}{2}\delta_{\alpha\beta} \partial^{\sigma}\hat{\phi}(x)\partial_{\sigma}\hat{\phi}(x) \, .
\end{equation}
We can write it in the following point-split form 
\begin{equation}
\hat{T}_{\alpha \beta}(x) = \lim_{y \to x} P_{\alpha \beta}(x,y)\big(\hat{\phi}(x)\hat{\phi}(y)\big) \, ,
\end{equation}
where 
\begin{equation}
P_{\alpha \beta}(x,y) = \partial_{\alpha}^{(x)}\partial_{\beta}^{(y)}\ - \frac{1}{2}\delta_{\alpha \beta} \partial^{\sigma (x)}\partial_{\sigma}^{(y)} \, .
\end{equation}
The superscripts $(x)$ and $(y)$ have been put to denote that the corresponding operators act only the field with the same spacetime argument in the point-split form.  \\
Using the above point-split form of the stress energy tensor and taking $\ket{\psi}$ to be a vacuum state $\ket{0}$, we can write the stress energy correlator as follows
\begin{equation}
\bra{0}\hat{T}_{\alpha\beta}(x)\hat{T}_{\gamma\delta}(x')\ket{0} = \lim_{y \to x}\lim_{y' \to x'}P_{\alpha\beta}(x,y) P_{\gamma \beta}(x',y')\bra{0}\hat{\phi}(x)\hat{\phi}(y)\hat{\phi}(x')\hat{\phi}(y')\ket{0} \, .
\end{equation}
The four point correlator of the field operators can be written as a sum of products of Wightman functions using the Wick's theorem \cite{Dhanuka:2020yxp}. In fact, we obtain 
\begin{equation}
\bra{0}\hat{T}_{\alpha\beta}(x)\hat{T}_{\gamma\delta}(x')\ket{0} = 2\lim_{y \to x}\lim_{y' \to x'}P_{\alpha\beta}(x,y) P_{\gamma \beta}(x',y') ( G(x,x')G(y,y')) \, ,
\end{equation}
where we have dropped the UV divergent $G(x,y)G(x',y')$ term on the account of its contribution vanishing in the computation of  the UdW response rates as has been argued in \cite{Padmanabhan:1987rq}. 
Therefore, we have derivative operators acting on a product of Wightman functions. Thus, we have a generalized version of the derivative coupling that we encountered in the previous section. Earlier, we had only time derivatives of one Wightman function determining the transition probability and the response rate but here in this case, we have all types of derivative operators acting on product of two Wightman functions. Thus, if there is any spacetime independent infrared divergent term appearing in the Wightman function, then it is cured by the derivative operators. However, the spacetime dependent infrared divergent terms still make contributions to the response rate and even dominate the response rates. In order to demonstrate this, let us expand the above expression i.e., 
\begin{multline}
\bra{0}\hat{T}_{\alpha\beta}(x)\hat{T}_{\gamma\delta}(x')\ket{0} = \Bigg(\partial_{\beta}\partial_{\gamma}'G(x,x')\partial_{\alpha}\partial_{\delta}'G(x,x') + \partial_{\beta}\partial_{\delta}'G(x,x')\partial_{\alpha} \partial_{\gamma}'G(x,x') \\- \eta_{\gamma \delta}\eta^{\rho \sigma}\partial_{\alpha}\partial_{\rho}'G(x,x')\partial_{\beta}\partial_{\sigma}'G(x,x') \\ -\eta_{\alpha \beta}\eta^{\rho \sigma}\partial_{\rho}\partial_{\gamma}'G(x,x')\partial_{\sigma}\partial_{\delta}'G(x,x') +  \frac{1}{2}\eta_{\alpha \beta}\eta^{\mu \nu}\eta_{\gamma \delta}\eta^{\rho \sigma}\partial_{\mu}\partial_{\rho}'G(x,x')\partial_{\nu}\partial_{\sigma}'G(x,x')  \Bigg) \, .
\end{multline}

If we specialize to a class of such detectors with internal states for which only $\langle \Omega|\hat{\mu}^{00}|0\rangle_D $ survives while all other components cross elements are zero , what we have effectively is the energy coupled UdW detectors in which the response rate depends upon the energy-energy correlator.
For such detectors, the relevant correlator is given by 
\begin{multline}
\bra{0}\hat{T}_{00}(x)\hat{T}_{00}(x')\ket{0} = \frac{1}{2}\Bigg(\partial_{0}\partial_{0}'G(x,x')\partial_{0}\partial_{0}'G(x,x')  + \partial_{0}\partial_{k}'G(x,x')\partial_{0}\partial_{k}'G(x,x') \\ + \partial_{k}\partial_{0}'G(x,x')\partial_{k}\partial_{0}'G(x,x') +  \delta^{kp}\delta^{ln}\partial_{k}\partial_{l}'G(x,x')\partial_{p}\partial_{n}'G(x,x')  \Bigg) \, .
\end{multline}
For massless scalar fields in nearly matter dominated spacetimes, the Wightman function is given by Eq.~\eqref{522},
from which we see that, since the infrared divergent term has only time dependent factors, the spatial derivatives remove the $1/\delta$ term and the only term contributing $1/\delta$ is the all time derivatives term. Therefore, we obtain that 
\begin{equation}
\bra{0}\hat{T}_{00}(x)\hat{T}_{00}(x')\ket{0} = \frac{1}{\delta^2} \frac{1}{128\pi^4 H^8 } \frac{81}{(\eta_{1}\eta_{2})^8} + O(\delta^{-1})\, .
\end{equation}
Thus, we see that the time dependent infrared divergent term survives the derivative operators and it contributes dominantly to the response rate. When $\delta$ is taken very close to zero, then the above $1/(\delta)^2$ term will dominate the response rate. In case of nearly massless scalar fields in de Sitter spacetime, we need to be a bit careful as there would be a corresponding mass term in the stress energy operator expression which would not have any derivative operators in it. The additional mass term in the stress energy tensor also cures the infrared divergent term as this term would multiply the inverse mass square term of the Wightman function in Eq.~\eqref{521} and remove the divergent characteristic arising because of the massless limit.  Thus, we conclude that, even though the technical details of stress energy coupled UdW detectors differ from the derivatively coupled UdW detectors of the previous section, the remarks made in the previous section regarding the contribution of infrared divergent terms to the response of derivatively coupled UdW detectors more or less apply to the present case of stress energy coupled UdW detectors.   \\
\subsection{Coupling of Hydrogen atoms with gravitational waves}\label{hydratom}
\begin{figure}[h]
\centering
\includegraphics[height = 8cm]{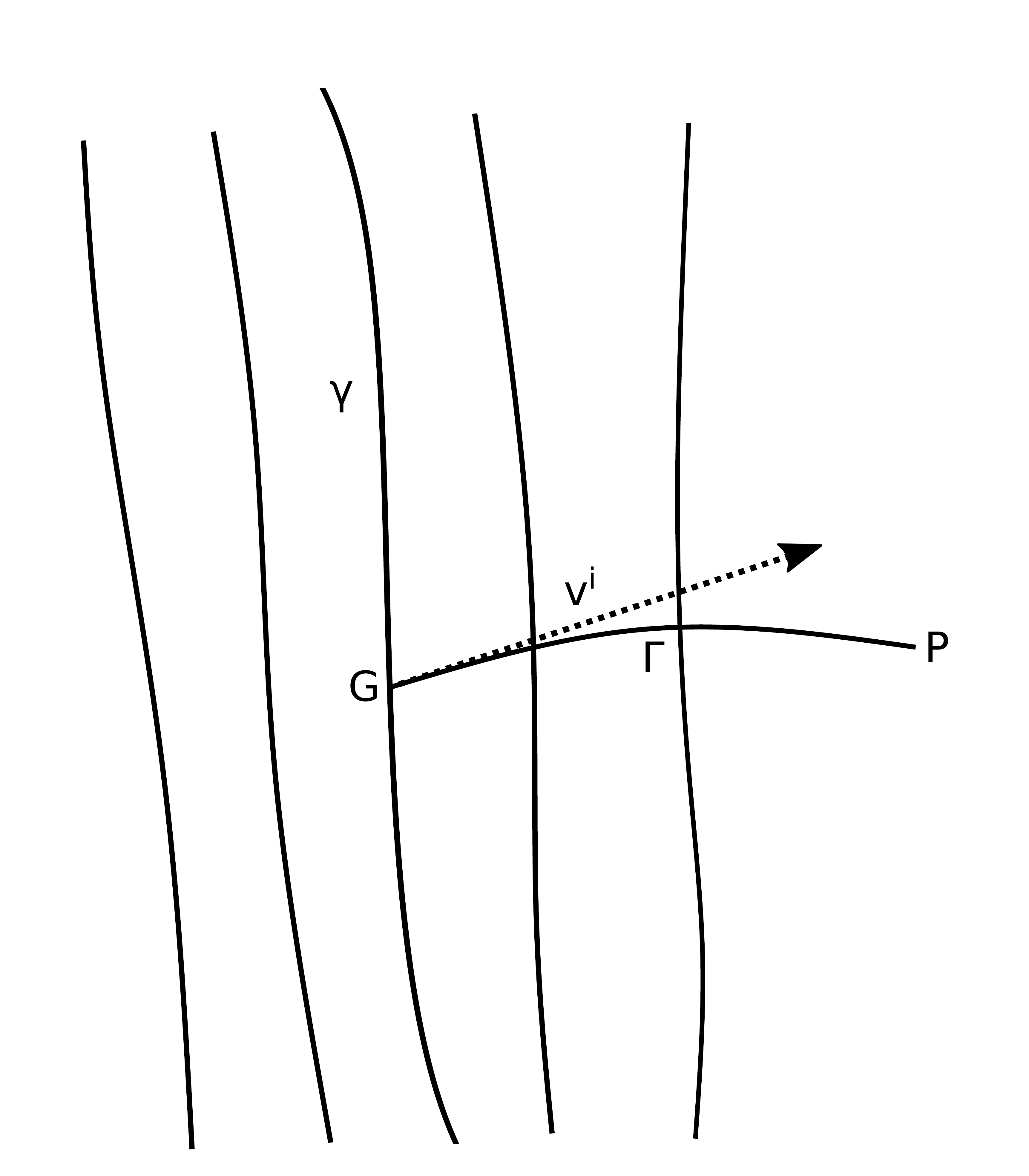}
\caption{This figure captures the construction of Fermi normal coordinates. Here, $\gamma$ is some time-like geodesic about which we construct FNCs. P is some point in the spacetime which intersects $\gamma$ orthogonally at point G via the unique spacelike geodesic $\Gamma$. $\text{v}^{i}$ are the components of the tangent vector to $\Gamma$ at point G.}\label{FNC_coor}
\end{figure}

\noindent This section explores the application of the above discussed results to the dynamics of an atom in a general curved spacetime with spacetime metric $g_{\mu\nu}$. To that end, we follow the treatment given in \cite{Parker:1980kw} which assumes that the center of mass of the atom moves along a classical time-like geodesic. To proceed further, one then builds Fermi normal coordinates (FNCs) around this `central' geodesic of the center of mass (see Fig. \ref{FNC_coor}). In FNCs,  for a point lying on the central geodesic, the time coordinate is taken to be the proper time along the central geodesic and the spatial coordinates are taken to be zero. For a spacetime point, $P$, lying off the central geodesic, the time coordinate is taken to be the proper time of point $G$, along the central geodesic at which the unique spacelike geodesic from the point, $P$, intersects the central geodesic orthogonally.  The spatial coordinates for point $P$, $x^{i} = s v^{i}$, where $s$ is the proper time along the unique spacelike geodesic from $G$ to $P$ and $v^{i}$ is the tangent to this geodesic at point $G$. \\
\noindent Using the above contruction, the internal structure of the atom, in the non-relativistic limit, is shown (see Appendix \ref{fermiCoor} ) to be governed by the following Schrodinger equation form \cite{Parker:1980kw}\begin{equation}\label{542}
\Big(i\frac{\partial}{\partial t} -m\Big)\psi = \Big(-\frac{1}{2m}\nabla^2- \frac{\zeta}{r} + \frac{1}{2}m R_{0l0m}x^{l}x^{m}\Big)\psi \, ,
\end{equation}
where $R_{0l0m}$ are the Riemann tensor components expressing the curvature induced corrections to the flat spacetime Schrodinger equation with the central Coulomb potential of the nucleus. The curvature corrections are considered only upto 2nd order in FNCs. The Riemann tensor components are evaluated on the central geodesic in FNCs and can be related to the Riemann tensor components, $R^{arbitrary}_{\mu\nu\gamma\delta}$, in any arbitrary coordinate system by the relation
\begin{equation}\label{543}
R^{FNC}_{abcd} = R^{arbitrary}_{\mu\nu\gamma\delta}\vec{e}^{\mu}_{a}\vec{e}^{\nu}_{b}\vec{e}^{\gamma}_{c}\vec{e}^{\delta}_{d} \, ,
\end{equation}
where $\vec{e}^{\mu}_{a}$ are a set of orthonormal basis parallel transported along the central timelike geodesic. The $\vec{e}^{\mu}_{0}$ is the tangent vector field along the central geodesic. \\

\noindent In the following, the curvature effects of FRW spacetimes are considered on an atom. After that by introducing metric perturbations over FRW spacetimes, we discuss if there are any observable effects of these perturbations on the atom. 


\subsubsection{FRW Spacetimes With No Perturbation}
\noindent For this analysis, the center of mass of the atom is taken to move along comoving trajectories in a flat FRW spacetime (with scale factor being $a(\eta)$). 
 For comoving observers, the spatial coordinates are fixed i.e., $x^{\mu}(t) = (\eta(t), c^{i})$ and the tangent vector field is given by 
\begin{equation}\label{546}
\frac{dx^{\mu}}{dt} = \Big(\frac{1}{a},0\Big).
\end{equation}
A set of orthonormal basis which are parallel transported along these comoving geodesics can be taken as follows
\begin{equation}\label{547}
\vec{e}^{\mu}_0 = \frac{1}{a}(1,0,0,0),\ \ \vec{e}^{\mu}_1 = \frac{1}{a}(0,1,0,0), \ \ \vec{e}^{\mu}_2 = \frac{1}{a}(0,0,1,0),\ \ \vec{e}^{\mu}_3 = \frac{1}{a}(0,0,0,1).
\end{equation}
The relevant Riemann tensor components in FNCs, using the relation 
\begin{equation}\label{548}
R^{FNC}_{0l0m} = R^{Con}_{\mu\nu\gamma\delta}\vec{e}^{\mu}_{0}\vec{e}^{\nu}_{l}\vec{e}^{\gamma}_{0}\vec{e}^{\delta}_{m} \, , 
\end{equation}
are given by the relation $R^{FNC}_{0l0m} = R^{Con}_{0l0m}/a^4$. Using the form of the Riemann tensor components in conformal coordinates i.e., $R^{Con}_{0l0m} = -\delta_{lm}(aa'' - a'^2)$, one finds that
\begin{equation}\label{549}
R^{FNC}_{0l0m} = -\delta_{lm}\frac{1}{a^4}(aa'' - a'^2) \, ,
\end{equation}
where $'$ denotes a derivative with respect to conformal time. Thus, the interaction Hamiltonian is given by 
\begin{equation}\label{550}
H_{I} = - \frac{m}{2}\frac{\ddot{a}}{a}r^2 \, ,
\end{equation}
where $\dot{}$ denotes a derivative with respect to comoving time coordinate. \\

\noindent With the above interaction Hamiltonian, the transition probability, upto to first order in perturbation theory, for the atom to go from some state $\psi_{nlm}$ to $\psi_{n'l'm'}$ under the comsological expansion of FRW spacetimes is given by
\begin{equation}\label{561}
P_{\psi_{nlm}\to \psi_{n'l'm'}} = \frac{m^2}{4}|\bra {\psi_{n'l'm'}}r^2\ket{\psi_{nlm}}|^2   \int^{\eta_{f}}_{\eta_{i}}d\eta_{1}\int^{\eta_{f}}_{\eta_{i}}d\eta_{2}e^{-i\Omega(t(\eta_{1})-t(\eta_{2}))} \frac{1}{a_{1}a_{2}}\Big(\frac{a_{1}''}{a_{1}} - \frac{a_{1}'^{2}}{a_{1}^{2}}\Big)\Big(\frac{a_{2}''}{a_{2}} - \frac{a_{2}'^{2}}{a_{2}^{2}}\Big) \, ,
\end{equation}
where $\Omega = E_{n'l'}-E_{nl}$ is the energy difference between the two states.\\ 
Because of the isotropic form of the interaction Hamiltonian, the transition probability is found to be zero, at least upto to the order under consideration, for the cases in which the atom makes a transition between states with different spherical harmonics. More precisely, we see that 
\begin{equation}\label{562}
\bra {\psi_{n'l'm'}}r^2\ket{\psi_{nlm}} = \int dr\ r^4 R_{n'l'}^{*}(r)R_{nl}(r) \int\int sin\theta\ d\theta\ d\phi\  Y^{m'*}_{l'}(\theta,\phi)\ Y^{m}_{l}(\theta,\phi) \, ,
\end{equation}
and because of the orthonormality of the spherical harmonics, the above integral vanishes for mismatching $(m,l)$ and $(m',l')$. Selection rules corresponding to the detector terms like $\hat{x}^{i}\hat{x}^{k}$ are discussed in Appendix \ref{flathydr}. \\
This is something that is expected because the cosmological expansion in FRW spacetimes respects the spherical symmetry of the spatial slices for all times. Other important thing to notice is that these transitions are caused by the classical expansion of the FRW spacetimes, we have not assumed anything quantum about the spacetime. Therefore, if it is found that the quantized tensorial metric perturbations over FRW spacetimes can cause transitions between states of the hydrogen atoms which have different spherical harmonics, then the observations of these transitions with non-trivial change of spherical harmonics would be a purely quantum effect.

\subsubsection{Perturbed FRW Spacetimes}
\noindent Now let us consider the case of perturbed FRW spacetimes i.e., 
\begin{equation}\label{551}
ds^2 = a^2(\eta)(\eta_{\mu\nu} + h_{\mu\nu})dx^{\mu} dx^{\nu} \, .
\end{equation}
The perturbation in the FRW metric results in the perturbations to the comoving geodesics of the FRW spacetimes. In fact, it can be shown (see \cite{Dai:2015rda}) that the set of parallel transported orthonormal basis upto first order in $h$ is given by 
\begin{eqnarray}\label{552}
\vec{e}^{\mu}_0 &=& \frac{1}{a}(1 + \frac{h_{00}}{2},V_{0i}) \, ,\\ \vec{e}^{\mu}_i &=& \frac{1}{a}(V_{0i} + h_{0i},\delta^{j}_{i}-\frac{h^{j}_{i}}{2} + \frac{1}{2}\epsilon^{jk}_{i}\omega_{k}), 
\end{eqnarray}
where $V_{0i}, \omega_{k}$ denote perturbations to the comoving geodesics along with the other factors of $h_{\mu\nu}$ and they are given by
\begin{eqnarray}\label{553}
 V_{0i}' + \frac{a'}{a}V_{0i}&=& \frac{1}{2}\partial_{i}h_{00}-h_{0i}'-\frac{a'}{a}h_{0i} \, , \\ 
 \omega_{k}'&=& -\frac{1}{2}\epsilon_{k}^{ij}(\partial_{i}h_{oj}-\partial_{j}h_{oi}) \, .
\end{eqnarray}
Though one can study the effect of all of scalar, vector and tensor perturbations, here we consider the case of gravitational waves only for which $h_{00}=h_{0i}=0$ and $h_{ij}$ is such that $h_{ij}\delta^{ij} = 0$ and $\delta^{ki}\partial_{k}h_{ij} = 0$.  This implies that $V_{0i} = \omega_{k} =0$ and the perturbed vierbeins are given by 
 \begin{eqnarray}\label{554}
\vec{e}^{\mu}_0 &=& \frac{1}{a}(1 ,0) \, ,\\ \vec{e}^{\mu}_i &=& \frac{1}{a}(0,\delta^{j}_{i}-\frac{h^{j}_{i}}{2} ) \, .
\end{eqnarray}
Thus, the tangent vector field to the perturbed geodesic is not affected upto first order in $h$ and hence the proper time for the geodesic is just the cosmic time upto first order and the spatial coordinates are also fixed upto first order in $h$ i.e., $x^{\mu}(t) = (\eta(t), c^{i})$ upto first order in $h$.\\ \\
Now using these orthonormal basis vector fields, we can convert the Reimann tensor from conformal coordinate system to the Fermi normal coordinates (FNCs) by the relation 
\begin{equation}\label{555}
R^{FNC}_{abcd} = R^{Con}_{\mu\nu\gamma\delta}\vec{e}^{\mu}_{a}\vec{e}^{\nu}_{b}\vec{e}^{\gamma}_{c}\vec{e}^{\delta}_{d} \, .
\end{equation}
Particularly, 
\begin{eqnarray}\label{556}
R^{FNC}_{0l0m} &=& R^{Con}_{\mu\nu\gamma\delta}\vec{e}^{\mu}_{0}\vec{e}^{\nu}_{l}\vec{e}^{\gamma}_{0}\vec{e}^{\delta}_{m} \nonumber \\
&=& \frac{1}{a^4}R^{Con}_{0k0p}\big(\delta^{k}_{l}-\frac{h^{k}_{l}}{2}\big)\big(\delta^{p}_{m}-\frac{h^{p}_{m}}{2}\big) \, .
\end{eqnarray}
The Riemann tensor components relevant for our purposes, in conformal coordinates, are given (can be obtained from expressions given in \cite{Mukhanov:1990me, Bartolo:2004if}) by 
\begin{equation}\label{557}
R^{Con}_{0l0m} = -\delta_{lm}(aa'' - a'^2) - (aa''-a'^2)h_{lm}-\frac{aa'}{2}h_{lm}^{'} -\frac{a^2}{2} h_{lm}^{''} \, .
\end{equation}
Using this, the relevant Riemann tensor components in the FNCs are given by
\begin{eqnarray}\label{558}
R^{FNC}_{0l0m} 
&=& \frac{1}{a^4}\Big(-\delta_{kp}(aa'' - a'^2) - (aa''-a'^2)h_{kp}-\frac{aa'}{2}h_{kp}^{'} -\frac{a^2}{2} h_{kp}^{''}\Big) \Big(\delta^{k}_{l}-\frac{h^{k}_{l}}{2}\Big)\Big(\delta^{p}_{m}-\frac{h^{p}_{m}}{2}\Big) \nonumber \\
& = & \frac{1}{a^4}\Big(-\delta_{lm}(aa'' - a'^2) -\frac{aa'}{2}h_{lm}^{'} -\frac{a^2}{2} h_{lm}^{''}\Big)  + O(h^2) \, .
\end{eqnarray}
Thus the above expression for the relevant Riemann components implies that the interaction Hamiltonian is 
\begin{eqnarray}\label{559}
H_{I} &=& \frac{m}{2} \frac{1}{a^4}\Big(-\delta_{lm}(aa'' - a'^2) -\frac{aa'}{2}h_{lm}^{'} -\frac{a^2}{2} h_{lm}^{''}\Big)x^{l}x^{m} \nonumber \\
&=& \frac{m}{2} \Big(-\delta_{lm}\frac{\ddot{a}}{a} -\frac{\dot{a}}{a}\dot{h}_{lm} -\frac{1}{2} \ddot{h}_{lm}\Big)x^{l}x^{m} = \frac{m}{2}H_{lm}x^{l}x^{m} \, .
\end{eqnarray}
Here $'$ again denotes a derivative with respect to conformal time, $\eta$, and $\dot{}$ denotes a derivative with respect to comoving time, $t$. Here $H_{lm} = \Big(-\delta_{lm}(\ddot{a}/a) -(\dot{a}/a)\dot{h}_{lm} -(1/2) \ddot{h}_{lm}\Big).$ \\ \\
Now let us consider the case in which an atom makes a transition from $\psi_{nlm}$ to $\psi_{n'l'm'}$ while the tensor perturbation starts in an initial vacuum state and is allowed to go to any arbitrary final state. The transition probability for this case is given by 
\begin{multline}\label{564}
P_{\psi_{nlm}\to \psi_{n'l'm'}} = \frac{m^2}{4}\bra {\psi_{n'l'm'}}\hat{x}^{i}\hat{x}^{j}\ket{\psi_{nlm}}^{*}\bra {\psi_{n'l'm'}}\hat{x}^{p}\hat{x}^{k}\ket{\psi_{nlm}}  \\ \int^{\eta_{f}}_{\eta_{i}}d\eta_{1} \int^{\eta_{f}}_{\eta_{i}}d\eta_{2}e^{-i\Omega(t({\eta_{1}})-t(\eta_{2}))}a(\eta_{1})a(\eta_{2})\bra{0} \hat{H}_{ij}(\vec{c},\eta_{1})\hat{H}_{pk}(\vec{c},\eta_{2})\ket{0} \, ,
\end{multline}
where $\Omega = E_{n'l'}-E_{nl}$ is the energy difference between the two atomic states and $\vec{c}$ represents the fixed spatial coordinates for the comoving trajectory along which the atom is moving. In the absence of $h_{\mu\nu}$, we recover the case of purely classical expansion of FRW spacetimes of the previous subsection and  hence cannot have transitions between eigenfunctions corresponding to different spherical harmonics. However, with gravitational waves, we expect to have transitions with non-trivial change in spherical harmonics.\\\\
The tensor perturbations in FRW spacetimes satisfy the following equation of motion \cite{Weinberg:2008zzc}
\begin{equation}\label{565}
h_{lm}^{''} + 2\frac{a'}{a}h_{lm}^{'} -\nabla^2 h_{lm} =0 \, ,
\end{equation}
which is just the same equation as that of a massless scalar field in FRW spacetimes. Using the fact that there are only two independent polarization states of gravitational waves in an FRW spacetime and they follow the same equation of motion as that of a massless scalar field evolving on the same FRW spacetime, one can say that the dynamics of tensor perturbations is equivalent to two massless scalar fields over the considered FRW spacetime. \\
Using this, the $H_{lm}$ can be written as 
\begin{equation}\label{566}
H_{lm} = -\frac{\delta_{lm}}{a^4}(aa'' - a'^2) + \frac{1}{a^4}\Big(\frac{aa'}{2}\frac{\partial}{\partial\eta} -\frac{a^2}{2} \nabla^{2}_{\vec{c}}\Big)h_{lm}(\eta,\vec{c}) \, .
\end{equation}
The quantized tensor perturbations can be expanded in terms of mode functions as 
\begin{equation}\label{567}
\hat{h}_{ij}(\vec{c},\eta) = \sum_{\lambda = +,\times}\int d^3\vec{q}\ e_{ij}(\hat{q}, \lambda) \Big(e^{i\vec{q}.\vec{c}}h_{q}(\eta)\hat{b}_{\vec{q},\lambda} + e^{-i\vec{q}.\vec{c}}h^{*}_{q}(\eta)\hat{b}^{\dagger}_{\vec{q},\lambda}\Big) \, ,
\end{equation}
where $\lambda = +, \times$ refer to two polarization states and $\vec{c}$ is the constant spatial vector for the considered comoving trajectories. The $e_{ij}(\vec{q},\lambda)'s$ satisfy \cite{Weinberg:2008zzc}
\begin{multline}\label{568}
\sum_{\lambda = +,\times}e_{ij}(\hat{q},\lambda)e_{kl}(\hat{q},\lambda) = \delta_{ik}\delta_{jl} + \delta_{il}\delta_{jk} - \delta_{ij}\delta_{kl} + \delta_{ij}\hat{q}_{k}\hat{q}_{l} + \delta_{kl}\hat{q}_{i}\hat{q}_{j} \\ -\delta_{ik}\hat{q}_{j}\hat{q}_{l}-\delta_{il}\hat{q}_{j}\hat{q}_{k}-\delta_{jk}\hat{q}_{i}\hat{q}_{l}-\delta_{jl}\hat{q}_{i}\hat{q}_{k} + \hat{q}_{i}\hat{q}_{j}\hat{q}_{k}\hat{q}_{l} \, .
\end{multline}
Also, $\hat{b}_{\vec{q},\lambda}$ and $\hat{b}^{\dagger}_{\vec{q},\lambda}$ are the annihilation and creation operators for a state with wave-vector, $\vec{q}$, and polarization, $\lambda$. The time dependent part, $h_{q}(\eta)$, of the mode functions satisfy 
\begin{equation}\label{569}
h_{q}^{''}(\eta) + 2\frac{a'}{a}h_{q}^{'}(\eta) + q^2 h_{q}(\eta) =0 \,  ,
\end{equation}
and hence the time evolution is independent of the direction of wave-vector and polarization state. \\ 
Let us now consider a transition between states with different spherical harmonics. For these transitions, many terms drop out. In particular, the term proportional to $\delta_{lm}$ in $H_{lm}$ drops out and we see that, using Eq. (\ref{566}), the vacuum expectation of the product of gravitational fields appearing in the formula for transition probability is given by
\begin{multline}\label{570}
\lim_{\vec{c}_1 \to \vec{c}_2}\bra{0} \hat{H}_{ij}(\vec{c}_{1},\eta_{1})\hat{H}_{pk}(\vec{c}_{2},\eta_{2})\ket{0} \\ = \lim_{\vec{c}_1 \to \vec{c}_2}\frac{1}{a_{1}^{4}}\Big(\frac{a_{1}a_{1}^{'}}{2}\frac{\partial}{\partial\eta_{1}} -\frac{a_{1}^{2}}{2} \nabla^{2}_{\vec{c}_{1}}\Big)\frac{1}{a_{2}^{4}}\Big(\frac{a_{2}a_{2}^{'}}{2}\frac{\partial}{\partial\eta_{2}} -\frac{a_{2}^2}{2} \nabla^{2}_{\vec{c}_{2}}\Big)\bra{0} \hat{h}_{ij}(\vec{c}_{1},\eta_1)\hat{h}_{pk}(\vec{c}_{2},\eta_2)\ket{0} \, .
\end{multline}

\noindent Using the Fourier expansion of gravitational waves i.e., Eq. (\ref{567}) and the commutation relations between creation and annihilation operators, one has  
\begin{eqnarray}\label{571}
\bra{0} \hat{h}_{ij}(\vec{c}_{1},\eta_1)\hat{h}_{pk}(\vec{c}_{2},\eta_2)\ket{0} &=& \sum_{\lambda = +,\times}\int d^3\vec{q}\ e_{ij}(\hat{q}, \lambda) e_{pk}(\hat{q}, \lambda)e^{i\vec{q}.(\vec{c}_{1}-\vec{c}_{2})}h_{q}(\eta_{1})h^{*}_{q}(\eta_{2}) \nonumber \\
&=& \int d^3\vec{q}\ \Big(\sum_{\lambda = +,\times}e_{ij}(\hat{q}, \lambda) e_{pk}(\hat{q}, \lambda)\Big)e^{i\vec{q}.(\vec{c}_{1}-\vec{c}_{2})}h_{q}(\eta_{1})h^{*}_{q}(\eta_{2}) \, .
\end{eqnarray}
Using Eq. (\ref{568}) and replacing every factor of $q_{i}$ in it by a partial derivative with respect to spatial coordinates outside the integral sign, the above Wightman function of gravitational waves can be written as a sum of products of  Kronecker delta's and spatial partial derivatives acting on the Wightman function of some scalar field i.e., 
\begin{multline}\label{572}
\bra{0} \hat{h}_{ij}(\vec{c}_{1},\eta_1)\hat{h}_{pk}(\vec{c}_{2},\eta_2)\ket{0} = \Big(\delta_{ip}\delta_{jk} + \delta_{ik}\delta_{jp} - \delta_{ij}\delta_{pk} + \delta_{ij}\frac{\partial_{\vec{c}_{1p}}\partial_{\vec{c}_{1k}}}{\nabla_{\vec{c}_1}^{2}} + \delta_{pk}\frac{\partial_{\vec{c}_{1i}}\partial_{\vec{c}_{1j}}}{\nabla_{\vec{c}_1}^{2}}  -\delta_{ip}\frac{\partial_{\vec{c}_{1j}}\partial_{\vec{c}_{1k}}}{\nabla_{\vec{c}_1}^{2}} \\ - \delta_{ik}\frac{\partial_{\vec{c}_{1j}}\partial_{\vec{c}_{1p}}}{\nabla_{\vec{c}_1}^{2}}-\delta_{jp}\frac{\partial_{\vec{c}_{1i}}\partial_{\vec{c}_{1k}}}{\nabla_{\vec{c}_1}^{2}}-\delta_{jk}\frac{\partial_{\vec{c}_{1i}}\partial_{\vec{c}_{1p}}}{\nabla_{\vec{c}_1}^{2}} + \frac{\partial_{\vec{c}_{1i}}\partial_{\vec{c}_{1j}}\partial_{\vec{c}_{1p}}\partial_{\vec{c}_{1k}}}{\nabla_{\vec{c}_1}^{2}\nabla_{\vec{c}_1}^{2}}\Big)\int d^3\vec{q} e^{i\vec{q}.(\vec{c}_{1}-\vec{c}_{2})}h_{q}(\eta_{1})h^{*}_{q}(\eta_{2}) \, .
\end{multline}
Looking at the Eqs. (\ref{564}), (\ref{570}) and (\ref{572}), we notice that the coupling of gravitational waves with atoms has a structure similar to that of UdW detectors which are derivatively coupled with quantum fields Eq. (\ref{526}). 
In this case, derivatives with respect to spatial coordinates also appear in addition to derivative with respect to time coordinate. Finally, using the Eqs. (\ref{564}), (\ref{570}) and (\ref{572}), the transition probability between states with different spherical harmonics is given by 
\begin{multline}\label{573}
P_{\psi_{nlm}\to \psi_{n'l'm'}} = \frac{m^2}{4}\bra {\psi_{n'l'm'}}\hat{x}^{i}\hat{x}^{j}\ket{\psi_{nlm}}^{*}\bra {\psi_{n'l'm'}}\hat{x}^{p}\hat{x}^{k}\ket{\psi_{nlm}}   \lim_{\vec{c}_1 \to \vec{c}_2}\int^{\eta_{f}}_{\eta_{i}}d\eta_{1}\int^{\eta_{f}}_{\eta_{i}}d\eta_{2}e^{-i\Omega(t(\eta_{1})-t(\eta_{2}))}\\ \frac{1}{a_{1}^{3}}\Big(\frac{a_{1}a_{1}^{'}}{2}\frac{\partial}{\partial\eta_{1}} -\frac{a_{1}^{2}}{2} \nabla^{2}_{\vec{c}_{1}}\Big)\frac{1}{a_{2}^{3}}\Big(\frac{a_{2}a_{2}^{'}}{2}\frac{\partial}{\partial\eta_{2}} -\frac{a_{2}^2}{2} \nabla^{2}_{\vec{c}_{2}}\Big)\Big(\delta_{ip}\delta_{jk} + \delta_{ik}\delta_{jp} - \delta_{ij}\delta_{pk} + \delta_{ij}\frac{\partial_{\vec{c}_{1p}}\partial_{\vec{c}_{1k}}}{\nabla_{\vec{c}_1}^{2}} + \delta_{pk}\frac{\partial_{\vec{c}_{1i}}\partial_{\vec{c}_{1j}}}{\nabla_{\vec{c}_1}^{2}}  -\delta_{ip}\frac{\partial_{\vec{c}_{1j}}\partial_{\vec{c}_{1k}}}{\nabla_{\vec{c}_1}^{2}} \\ - \delta_{ik}\frac{\partial_{\vec{c}_{1j}}\partial_{\vec{c}_{1p}}}{\nabla_{\vec{c}_1}^{2}}-\delta_{jp}\frac{\partial_{\vec{c}_{1i}}\partial_{\vec{c}_{1k}}}{\nabla_{\vec{c}_1}^{2}}-\delta_{jk}\frac{\partial_{\vec{c}_{1i}}\partial_{\vec{c}_{1p}}}{\nabla_{\vec{c}_1}^{2}} + \frac{\partial_{\vec{c}_{1i}}\partial_{\vec{c}_{1j}}\partial_{\vec{c}_{1p}}\partial_{\vec{c}_{1k}}}{\nabla_{\vec{c}_1}^{2}\nabla_{\vec{c}_1}^{2}}\Big)\int d^3\vec{q} e^{i\vec{q}.(\vec{c}_{1}-\vec{c}_{2})}h_{q}(\eta_{1})h^{*}_{q}(\eta_{2}) \, .
\end{multline}

\noindent Let us now look at some of the implications of this result. First, we notice that, since only the transitions between atomic states with different spherical harmonics are considered, the contribution to the transition probability from the classical expansion of the FRW backgrounds is not there and such transitions are, in this sense, \textit{exclusively the result of the quantum fluctuations of the metric perturbations}. To be able to say more from the above expression, the quantum state needs to be specified in which the tensor perturbations are placed. Since the dynamics of tensor perturbations over an FRW spacetime is equivalent to the dynamics of two massless scalar fields over the same FRW spacetime, we can make use of the properties of massless scalar fields in FRW spacetimes to analyse our case. For example, the integral in the above expression can be taken to be the Wightman function Eq. (\ref{510}) that was used for discussions in the sections \ref{UdW} and \ref{dUdW}. One can look at many aspects of the above formula with the mentioned Wightman function but one potentially important implication are the very fast transitions of electrons within the atomic states of the hydrogen atom when it passes through the phases of the universe which are nearly matter dominated. This can be immediately seen by recalling what we have studied about massless scalar fields in nearly matter dominated spacetimes which are derivatively coupled with UdW detectors. The fact that the $1/\delta$ term in the Wightman function Eq. (\ref{522}) has spacetime dependence and it does not vanish under time derivatives, implies that the $1/\delta$ term provides the dominant contribution to the response rate in the $\delta \to 0$ limit. This observation can have potentially important implications in late time era of the expansion of the universe, which we discuss in a related work.  

\section{Summary}\label{con}
\noindent This work has analysed the response rate of Unruh deWitt detectors which couple to quantum scalar fields in FRW spacetimes. We have looked at the case of both conventionally and derivatively coupled UdW detectors. In order to carry out this task, an equivalence \cite{Lochan:2018pzs} has been employed that exists between massless scalar fields in FRW spacetimes with massive scalar fields in de Sitter spacetime and the fields of FRW spacetimes have been placed in the Bunch Davies like vacuum of the corresponding massive scalar fields of de Sitter spacetime. We have also provided few examples i.e., the stress energy coupled UdW detectors and the interaction of hydrogen atom with gravitational waves in FRW spacetimes where a derivatively coupled UdW detector like interaction appears and one can carry over the analysis of derivatively coupled UdW detectors to these examples. The main results of this work can be summarized as follows:
\begin{enumerate}
\item{\textbf{Response rate of conventionally coupled UdW detectors :} First, we look at the coupling of a conventional UdW detector with massless scalar fields in FRW spacetimes and nearly massless scalar fields in de Sitter spacetime. It has been argued that for massless scalar fields in considered FRW spacetimes, the infinite time response rate increases with increasing $H$ for $q \in (-2,0)$ whereas it decreases with increasing $H$ for $q \in (0,1)$. We also consider the finite time response rate for these cases. For nearly massless scalar fields in de Sitter spacetime, the term which gives rise to the infrared divergence in the massless limit at the level of Wightman function manifests itself at the response rate level and provides the dominant contribution to the response rate in the massless limit. Since the massless fields in nearly matter dominated case are conformally related to the nearly massless scalar fields in de Sitter spacetime (by the equivalence between fields in FRW and de Sitter spacetimes), the infrared divergence of the de Sitter case is inherited by the nearly matter dominated spacetimes and because of this fact, the response rate of UdW detectors coupled with massless fields in nearly matter dominated spacetimes also possess an infrared divergent term which dominates the response rate in the spacetimes going to the matter dominated limit. We analyse the response rate of UdW detectors for some other FRW spacetimes also but the mentioned divergence occurs for the de Sitter and matter dominated spacetimes.   }
\item{\textbf{Response rate of derivatively coupled UdW detectors :} Next, we analyse the behaviour of the response rate of UdW detectors which are derivatively coupled to massless scalar fields in FRW spacetimes and nearly massless scalar fields in de Sitter spacetime. It is found that the infinite time response rate for this derivatively coupled case has the same dependence on H as for the conventional UdW case. We then take up the case of finite time response rate. Unlike in the case of conventional UdW detectors, the term which gives rise to infrared divergence in the massless limit for scalar fields in de Sitter spacetime does not contribute to the response rate. However, for massless scalar fields in nearly matter dominated spacetimes, the corresponding term contributes to the response rate and leads to faster and faster transition rates as the spacetimes approach to the matter dominated limit. As argued above, the reason for this is the fact that the response rate for derivatively coupled UdW detectors depends on the derivatives of the Wightman function and the infrared divergent term for de Sitter case does not have any spacetime dependence and hence it vanishes under the action of derivatives. Whereas for the nearly matter dominated spacetimes, the corresponding term has the spacetime dependence which survies the derivatives and contributes dominantly to the response rate. For this case, detectors of sufficiently small energy gap also behave as if they are put in a high temperature bath, to some extent. The response rate of these detectors for other FRW spacetimes do not show any divergences and hence, for the considered spacetimes, it is only the matter dominated spacetime for which the response rate shows infrared divergences.  }
\item{\textbf{Stress-energy tensor coupled UdW detectors:} One example where the derivative UdW like coupling appears is the case in which the UdW detector couples linearly with the stress energy operator, instead of the field operator. In this case, the transition probability (and hence the response rate) depends on the correlations of the stress energy operators along the trajectory of the spacetime. We argue that, for this case, because of the derivative operators present in the stress energy tensor, any spacetime independent infrared divergences would not contribute dominantly to the response rate just as has been found for the case of derivatively coupled UdW detectors . But the spacetime dependent infrared divergences of the matter dominated spacetime would survive the action of derivative operators of the stress energy tensor and provide the dominant contribution to the response rate.    }
\item{\textbf{Interaction of hydrogen atoms with gravitational waves :} Working in the leading order in FNCs, we consider the dynamics of hydrogen atoms in FRW spacetimes with and without metric perturbations and find that the interaction of hydrogen atoms with gravitational waves takes a form similar to derivatively coupled UdW detectors. From the above analysis for FRW spacetimes with no gravitational waves, it is seen that the classical expansion of the FRW spacetimes can cause the transitions of electrons in hydrogen atoms only between the atomic states which have same spherical harmonics. This result is expected as the expansion of an FRW spacetime respects the homogeneity and isotropy of the spatial slices of the spacetime. However, the quantized gravitational waves over an FRW background have been shown to be able to cause transitions also between the states which involve a non-trivial change of angular momentum quantum numbers. As mentioned before, the coupling of gravitational waves with hydrogen atoms has similarities with derivatively coupled UdW detectors, therefore the results obtained for derivatively coupled UdW detectors can be used to analyse the response rate of hydrogen atoms which are interacting with gravitational waves.  }
\end{enumerate}
From this study, we see that, for massless scalar fields in late time era particularly in nearly matter dominated spacetimes, the transitions within the internal states of UdW detectors for both conventional and derivatively coupled cases occur at very rapid rate and become larger and larger as the spacetime approach the matter dominated limit. We have also seen that there are physical systems which can capture these quantum effects. Previous studies have suggested quantum back reaction to become important even at late time phases of the evolution of our universe \cite{Dhanuka:2020yxp}. Hence, these potentially important quantum effects and probes during the late matter dominated epoch of the evolution of our universe should be scrutinized more carefully and we take up this task in a subsequent work. 
\section*{Acknowledgments}
\noindent AD would like to acknowledge the financial support from University Grants Commission, Government of India, in the form of Junior Research Fellowship (UGC-CSIR JRF/Dec-2016/510944). Research of KL is partially supported by the Department of Science and Technology (DST) of the Government of India through a research grant under INSPIRE Faculty Award (DST/INSPIRE/04/2016/000571). The authors thank S. Shankaranarayanan for providing useful comments on the manuscript. Authors also acknowledge Wolfram Mathematica which has been used to plot some of the figures. 

\newpage
\appendix
\section{Infinite time response rate of UdW detectors in FRW spacetimes}\label{infres}
\noindent In this appendix, we consider the infinite time response rate of conventionally and derivatively coupled UdW detectors in FRW spacetimes i.e., $a(\eta) = (H\eta)^{-q}$ with $q \in (-2,1)$. 
By making use of the dimensional analysis, we try to find out the dependence of the response rate on the energy gap between the detector's states as well as on the parameter $H$. 
\subsection{Conventionally coupled Unruh deWitt detector}
\noindent Let us consider the spacetimes with $q \in (-2,0)$ for which the cosmic time is related to the conformal time by the relation $t = \frac{H^{-q}\eta^{1-q}}{1-q}$ and $t \in (0, \infty)$ for $\eta \in (0, \infty)$. Using these relations and the formulae (\ref{504}),(\ref{510}), we find that the transition probability is given by 
\begin{equation}
    P_{0\rightarrow \Omega}=c^2|{}_{D}\braket{\Omega|\hat{\mu}(0)|0}_{D}|^2\int_{0}^{\infty} \int_{0}^{\infty} d\eta_{1} d\eta_{2}e^{i\frac{\Omega H^{-q}}{(1-q)}(\eta_{1}^{1-q}-\eta_{2}^{1-q})}(H^2\eta_{1}\eta_{2})^{-1}G^{dS}(y(x(\eta_{1}),x(\eta_{2}))) \, .
\end{equation}
Now defining a new variable $z = \Omega H^{-q}\eta^{1-q}$, we can pull out all the $\Omega$ and $H$ dependence out of the integral 
\begin{eqnarray}
\frac{P_{0\rightarrow \Omega}}{c^2|{}_{D}\braket{\Omega|\hat{\mu}(0)|0}_{D}|^2} &=& \int_{0}^{\infty} \int_{0}^{\infty} \frac{dz_{1} dz_{2}}{(1-q)^2}\frac{(z_{1}z_{2})^{\frac{q}{1-q}}}{(\Omega H^{-q})^{\frac{2}{1-q}}}e^{i\frac{(z_{1}-z_{2})}{(1-q)}}\frac{(\Omega H^{-q})^{\frac{2}{1-q}}}{H^2(z_{1}z_{2})^{\frac{1}{1-q}}}G^{dS}(y(x(\eta_{1}),x(\eta_{2}))) \nonumber \\ 
 &=& \int_{0}^{\infty} \int_{0}^{\infty} \frac{dz_{1} dz_{2}}{(1-q)^2}e^{-i\frac{(z_{1}-z_{2})}{(1-q)}}\frac{\Gamma\Big(\frac{3}{2} + \nu\Big)\Gamma\Big(\frac{3}{2} - \nu\Big) }{16\pi^2(z_{1}z_{2})}{}_2F_1\Big(\frac{3}{2} + \nu,\frac{3}{2} - \nu,2,1-\frac{y}{4}\Big)\, ,
\end{eqnarray}
where $y(z_{1},z_{2}) = -\frac{\big(z_{1}^{\frac{1}{1-q}}-z_{2}^{\frac{1}{1-q}} - i \epsilon (\Omega H^{-q})^{\frac{1}{1-q}}\big)^2}{(z_{1}z_{2})^{\frac{1}{1-q}}}$ for comoving observers. Since the only $\Omega$ and $H$ dependences in the integral are through the term $\epsilon (\Omega H^{-q})^{\frac{1}{1-q}}$ which go to zero in the $\epsilon \to 0$ limit, we find that the above integral does not depend upon $\Omega$ and $H$. However, the rate, say with respect to $\tilde{\eta} = \frac{\eta_{1} + \eta_{2}}{2}$ has the $\Omega$ and $H$ dependence of the following type  
\begin{eqnarray}
\frac{1}{c^2|{}_{D}\braket{\Omega|\hat{\mu}(0)|0}_{D}|^2}\frac{d P_{0\rightarrow \Omega}}{d\tilde{\eta}} & = &  \frac{1}{c^2|{}_{D}\braket{\Omega|\hat{\mu}(0)|0}_{D}|^2}\Big(\frac{\partial z_{1}}{\partial \tilde{\eta}}\frac{d}{dz_{1}} + \frac{\partial z_{2}}{\partial \tilde{\eta}}\frac{d}{dz_{2}}\Big) P_{0\rightarrow \Omega} \nonumber \\ 
& = & (\Omega H^{-q})^{\frac{1}{1-q}} (1-q)\Big(z_{1}^{\frac{q}{q-1}}\frac{d}{dz_{1}} + z_{2}^{\frac{q}{q-1}}\frac{d}{dz_{2}}\Big)\frac{P_{0\rightarrow \Omega}}{c^2|{}_{D}\braket{\Omega|\hat{\mu}(0)|0}_{D}|^2}  \nonumber \\&=& \frac{(\Omega H^{-q})^{\frac{1}{1-q}}}{(1-q)}\Bigg(z_{1}^{\frac{q}{q-1}}\int_{0}^{\infty} dz\Big[ e^{-i\frac{(z_{1}-z)}{(1-q)}}\frac{\Gamma\Big(\frac{3}{2} + \nu\Big)\Gamma\Big(\frac{3}{2} - \nu\Big) }{16\pi^2(z_{1}z)}   _2F_1\Big(\frac{3}{2} + \nu,\frac{3}{2} - \nu,2,1-\frac{y(z_{1},z)}{4}\Big)\Big]\nonumber \\ && + z_{2}^{\frac{q}{q-1}}\int_{0}^{\infty} dz\Big[ e^{-i\frac{(z-z_{2})}{(1-q)}}\frac{\Gamma\Big(\frac{3}{2} + \nu\Big)\Gamma\Big(\frac{3}{2} - \nu\Big) }{16\pi^2(z z_{2})}{} _2F_1\Big(\frac{3}{2} + \nu,\frac{3}{2} - \nu,2,1-\frac{y(z,z_{2})}{4}\Big)\Big]\Bigg)
 \nonumber \\&=& (\Omega H^{-q})^{\frac{1}{1-q}} f(q,z_{1},z_{2}) \propto (\Omega H^{-q})^{\frac{1}{1-q}} \, ,
\end{eqnarray} 
where the function $f(q,z_{1},z_{2})$ has no dependence on $\Omega$ and $H$ and it depends only on $q$, $z_{1}$ and $z_{2}$. The values of $z_{1}$ and $z_{2}$ combine to give the $\tilde{\eta}$ value at which the rate is being calculated. Similarly, for spacetimes with $q \in (0,1)$, the response rate can be shown to have the same $\Omega$ and $H$ dependences. 

\subsection{Derivatively coupled Unruh deWitt detector}
\noindent Let us again consider the spacetimes with $q \in (-2,0)$. In order to find the $\Omega$ and $H$ dependences of response rate for derivatively coupled cases, we make use of the following formulae for the infinite time transition probability for derivatively coupled UdW detectors  
\begin{equation}
    P_{0\rightarrow \Omega}=c^2|{}_{D}\braket{\Omega|\hat{\mu}(0)|0}_{D}|^2\int_{0}^{\infty} \int_{0}^{\infty} d\eta_{1} d\eta_{2}e^{-i\Omega(\tau(\eta_{1})-\tau(\eta_{2}))}\frac{d}{d\eta_{1}}\frac{d}{d\eta_{2}}G(x(\eta_{1}),x(\eta_{2})) \, .
\end{equation} 
Performing the same steps as in the previous subsection and using the formula (\ref{531}) for double derivatives of the Wightman function for FRW spacetimes, one obtains that the response rate for derivatively coupled UdW detectors has the following $\Omega$ and $H$ dependences
\begin{eqnarray}
\frac{1}{c^2|{}_{D}\braket{\Omega|\hat{\mu}(0)|0}_{D}|^2}\frac{dP_{0\rightarrow \Omega}}{d\tilde{\eta}} &\propto& \Omega^2(\Omega H^{-q})^{\frac{1}{1-q}} \, .
\end{eqnarray}
We obtain the same $\Omega$ and $H$ dependences of the response rate for spacetimes with $q \in (0,1)$.

\section{Derivatives of Wightman function}\label{dds} 
\noindent In this appendix, we express the double time derivative (appearing in the expression (\ref{528}) of the response rate for derivatively coupled UdW detectors) of the FRW Wightman function in terms of the derivatives of the de Sitter Wightman function by using the relation (\ref{510}) between the Wightman functions in the two settings i.e.,  
\begin{equation}
G^{FRW}(x_{1},x_{2}) = (H^2\eta_{1}\eta_{2})^{q-1}G^{dS}(y(x_1,x_2)) \, .
\end{equation}
Using the product rule of differentiation, we have  
\begin{eqnarray}
\frac{d}{d\eta_{1}}\frac{d}{d\eta_{2}}G^{FRW}(x(\eta_{1}),x(\eta_{2})) &=& (H^2\eta_{1}\eta_{2})^{q-1}\bigg[(q-1)^2\frac{G^{dS}}{\eta_{1}\eta_{2}} + (q-1)\frac{dG^{dS}}{dy}\Big(\frac{1}{\eta_{1}}\frac{dy}{d\eta_{2}} + \frac{1}{\eta_{2}}\frac{dy}{d\eta_{1}}\Big) \nonumber\\ & & + \Big(\frac{d^2G^{dS}}{dy^2}\frac{dy}{d\eta_{1}}\frac{dy}{d\eta_{2}} + \frac{dG^{dS}}{dy}\frac{d^2y}{d\eta_{1} d\eta_{2}}\Big)\bigg]  \, .
\end{eqnarray}
In conformal coordinates, the de Sitter invariant distance is given by
\begin{equation}
y = \frac{-(\eta_{1}-\eta_{2}-i\epsilon)^2 + (\Delta\vec{x})^2}{\eta_{1}\eta_{2}} \, .
\end{equation}
For comoving observers, we have 
\begin{eqnarray}
\frac{dy}{d\eta_{1}} &=& \frac{(\eta_{1}-\eta_{2}-i\epsilon)(-\eta_{1}-\eta_{2}-i\epsilon)}{\eta_{1}^{2}\eta_{2}} \, ,\\
\frac{dy}{d\eta_{2}} &=& \frac{(\eta_{1}-\eta_{2}-i\epsilon)(\eta_{1}+\eta_{2}-i\epsilon)}{\eta_{1}\eta_{2}^{2}} \, ,\\
\frac{d^2y}{d\eta_{1} d\eta_{2}} &=& \frac{-(\eta_{1}-\eta_{2}-i\epsilon)(\eta_{1}+\eta_{2}-i\epsilon) + 2(\eta_{1}-i\epsilon)\eta_{1}}{\eta_{1}^{2}\eta_{2}^{2}} \, ,\\
\frac{dy}{d\eta_{1}}\frac{dy}{d\eta_{2}} &=& \frac{y((\eta_{1} + \eta_{2})^2 + \epsilon^2)}{\eta_{1}^{2}\eta_{2}^{2}} \, .
\end{eqnarray}
Using these expressions, we see that 
\begin{eqnarray}
\frac{d}{d\eta_{1}}\frac{d}{d\eta_{2}}G^{FRW}(x(\eta_{1}),x(\eta_{2})) &=& (H^2\eta_{1}\eta_{2})^{q-1}\bigg[(q-1)^2\frac{G^{dS}}{\eta_{1}\eta_{2}} + (q-1)\frac{dG^{dS}}{dy}\Big(\frac{(\eta_{1}-\eta_{2}-i\epsilon)(-2i\epsilon)}{\eta_{1}^{2}\eta_{2}^{2}}\Big) \nonumber\\ & & + \frac{d^2G^{dS}}{dy^2}\frac{y((\eta_{1}+\eta_{2})^2 + \epsilon^2)}{\eta_{1}^{2}\eta_{2}^{2}} + \frac{dG^{dS}}{dy}\frac{(\eta_{1}^{2}+\eta_{2}^{2} + \epsilon^2)}{\eta_{1}^{2} \eta_{2}^{2}}\bigg]  \, .
\end{eqnarray}

\section{Covariant Dirac equation in FNCs}\label{fermiCoor}
\noindent In this appendix, we follow the treatment given in \cite{Parker:1980kw} and provide a very brief outline of how curvature effects are considered for atoms in curved spacetimes. We assume that the center of mass of an atom moves along a classical time-like geodesic whereas the internal structure of the atom is governed by the covariant Dirac equation for an electron in the presence of the electromagnetic potential of the nucleus i.e., we have the following equation for the internal structure of the atom 
\begin{equation}\label{538}
i\nabla_{0}\psi = \Big( - (g^{00})^{-1}\gamma^{0} m + i(g^{00})^{-1}\gamma^{0}\gamma^{i}\nabla_{i}\Big)\psi \, ,
\end{equation} 
where $\psi$ is a four component Dirac spinor and $\gamma^{\mu} = e^{\mu}_{a}\Gamma^{a}$ are the curved spacetime gamma matrices which are related to the flat spacetime gamma matrices, $\Gamma^{a}$, through the tetrad basis $e^{\mu}_{a}$. One also has $\{ \Gamma^{a},\Gamma^{b} \} = -2 \eta^{ab}$ and $e^{u}_{a}e^{\nu}_{b}g_{\mu\nu} = \eta_{ab}$. The covariant derivatives are given by 
\begin{equation}\label{539}
\nabla_{\mu} = \partial_{\mu}-\frac{1}{8}\omega^{ab}_{\,\  \mu}[\Gamma_a,\Gamma_{b}] - i q A_{\mu} \, ,
\end{equation} 
 where $\omega^{ab}_{\,\ \mu} = e^{a}_{\lambda}e^{\tau b}\Gamma^{\lambda}_{\tau \mu} - e^{\tau b}\partial_{\mu}e^{a}_{\tau}$. Here $A_{\mu}$ is the electromagnetic four potential which is determined by solving the curved spacetime Maxwell's equations in the presence of a point source at the nucleus. \\
To proceed further, we express the above equation in Fermi normal coordinates (FNCs) built around the `central' timelike geodesic (see Fig.~\ref{FNC_coor} and the discussion around it) of the center of mass of the atom. The spacetime metric upto to 2nd order in  these coordinates is  given as follows 
\begin{eqnarray}\label{540}
g_{00} &=& -1 - R_{0l0m}x^{l}x^{m} \, ,\\
g_{0i} &=& -\frac{2}{3}R_{0lim}x^{l}x^{m} \, ,\\
g_{ij} &=& \delta_{ij} - \frac{1}{3}R_{iljm}x^{l}x^{m} \, ,\\
g &=& -1 + \frac{1}{3}(R_{lm} - 2R_{0l0m})x^{l}x^{m} \, .
\end{eqnarray}
Similarly, one can express the inverse metric components, the tetrad bases and the Christoffel connections etc. in these coordinates upto to second order in the FNCs. For more details, refer to \cite{Parker:1980kw, Poisson:2009pwt}. \\
\noindent Now using the above form of the spacetime metric in the Dirac equation and the Maxwell's equations, one can show that the Dirac equation is given by 
\begin{equation}\label{541}
i \partial_{t}\psi = \Big(-i\alpha^{i}\partial_{i} + m \beta -\frac{\zeta}{r} + H_{I}\Big)\psi \, ,
\end{equation}
where $\beta, \alpha^{i}$ are the Dirac matrices and $H_{I}$ is the curvature induced perturbation to the flat spacetime Dirac equation in the central Coulomb potential $\frac{\zeta}{r}$. Here $\zeta = Ze^2$ where $e$ is the electron's charge and $Z$ is the number of protons in the nucleus. The expression for the interaction Hamiltonian, $H_{I}$, is given in \cite{Parker:1980kw}.\\
\noindent In the non-relativistic limit, the above Dirac equation can be shown to go to the following Schrodinger equation form \cite{Parker:1980kw}\begin{equation}\label{542}
\Big(i\frac{\partial}{\partial t} -m\Big)\psi = \Big(-\frac{1}{2m}\nabla^2- \frac{\zeta}{r} + \frac{1}{2}m R_{0l0m}x^{l}x^{m}\Big)\psi \, ,
\end{equation}
where $R_{0l0m}$ are the Riemann tensor components evaluated in FNCs along the central geodesic. In the above equation, the curvature induced perturbations are considered only upto 2nd order in FNCs. Here $\psi$ is now only one component function of space and time. 

\section{Flat Spacetime Hydrogen Atom and Selection Rules:}\label{flathydr}
\noindent We can solve for the energy eigenfunctions of the unperturbed flat spacetime hydrogen atom Hamiltonian. We find that they are given by \cite{Griffiths1}
\begin{equation}\label{544}
\psi_{nlm} = R_{nl}(r) Y^{m}_{l}(\theta,\phi) \, ,
\end{equation}
where $Y^{m}_{l}(\theta,\phi)$ are the spherical harmonics and $R_{nl}(r)$ are the radial part of the eigenfunctions. $R_{nl}(r)$ are given by 
\begin{equation}\label{545}
R_{nl}(r) = -\sqrt{\Big(\frac{2}{na_{0}}\Big)^3 \frac{(n-l-1)!}{2n((n + l)!)^3}}e^{-\frac{r}{na_{0}}}\Big(\frac{2r}{na_{0}}\Big)^{l} L^{2l + 1}_{n+l}\Big(\frac{2r}{na_{0}}\Big) \, ,
\end{equation}
where $L_{n+l}^{2l+1}$ are associated Laguerre polynomials and $a_{0} = \frac{1}{me^2}$. Here $n,l$ and $m$ are just the hydrogen atom quantum numbers. \\
Using the orthonormality of spherical harmonics and the properties of addition of angular momenta, we find that the selections rules for the transitions
\begin{equation*}
<n',l',m'|x^{i}|n,l,m> \, 
\end{equation*}
are as given in the table \ref{table1}.
\begin{table}[h]
\centering
\begin{tabular}{|cc|cc|}
\hline
\multicolumn{2}{|c|}{$x,y$}                                       & \multicolumn{2}{c|}{$z$}                               \\ \hline
\multicolumn{1}{|c|}{$\Delta l$}           & $\Delta m$           & \multicolumn{1}{c|}{$\Delta l$}           & $\Delta m$ \\ \hline
\multicolumn{1}{|c|}{$\pm$ 1} & $\pm$ 1 & \multicolumn{1}{c|}{$\pm$ 1} & 0          \\ \hline
\end{tabular}
\caption{Selection rules for transitions of the type $<n',l',m'|x^{i}|n,l,m>$. }\label{table1}
\end{table}\\
Using the above selection rules, we can find the selection rules for the transitions of the form 
\begin{equation*}
<n',l',m'|x^{i}x^{p}|n,l,m> \, 
\end{equation*}
which are given in the table \ref{table2}. 
\begin{table}[h]
\centering
\begin{tabular}{|cc|cc|cc|}
\hline
\multicolumn{2}{|c|}{$x^2, y^2, xy, yx$}      & \multicolumn{2}{c|}{$xz, yz$}              & \multicolumn{2}{c|}{$z^2$}                   \\ \hline
\multicolumn{1}{|c|}{$\Delta l$} & $\Delta m$ & \multicolumn{1}{c|}{$\Delta l$} & $\Delta m$ & \multicolumn{1}{c|}{$\Delta l$} & $\Delta m$ \\ \hline
\multicolumn{1}{|c|}{-2,0,2}     & -2,0,2     & \multicolumn{1}{c|}{-2,0,2}     & -1,1       & \multicolumn{1}{c|}{-2,0,2}     & 0          \\ \hline
\end{tabular}
\caption{Selection rules for transitions of the type $<n',l',m'|x^{i}x^{p}|n,l,m>$.}\label{table2}
\end{table}\\

\bibliographystyle{unsrt}
\bibliography{ref}

\begin{thebibliography}{10}

\bibitem{Birrell:1982ix}
N.~D. Birrell and P.~C.~W. Davies.
\newblock {\em {Quantum Fields in Curved Space}}.
\newblock Cambridge Monographs on Mathematical Physics. Cambridge Univ. Press,
  Cambridge, UK, 2 1984.

\bibitem{Unruh:1976db}
W.~G. Unruh.
\newblock {Notes on black hole evaporation}.
\newblock {\em Phys. Rev. D}, 14:870, 1976.

\bibitem{Crispino:2007eb}
Luis C.~B. Crispino, Atsushi Higuchi, and George E.~A. Matsas.
\newblock {The Unruh effect and its applications}.
\newblock {\em Rev. Mod. Phys.}, 80:787--838, 2008.

\bibitem{Svaiter:1992xt}
B.~F. Svaiter and N.~F. Svaiter.
\newblock {Inertial and noninertial particle detectors and vacuum
  fluctuations}.
\newblock {\em Phys. Rev. D}, 46:5267--5277, 1992.
\newblock [Erratum: Phys.Rev.D 47, 4802 (1993)].

\bibitem{Brown:2012pw}
Eric~G. Brown, Eduardo Martin-Martinez, Nicolas~C. Menicucci, and Robert~B.
  Mann.
\newblock {Detectors for probing relativistic quantum physics beyond
  perturbation theory}.
\newblock {\em Phys. Rev. D}, 87:084062, 2013.

\bibitem{Lin:2006jw}
Shih-Yuin Lin and B.~L. Hu.
\newblock {Backreaction and the Unruh effect: New insights from exact solutions
  of uniformly accelerated detectors}.
\newblock {\em Phys. Rev. D}, 76:064008, 2007.

\bibitem{Juarez-Aubry:2014jba}
Benito~A. Ju\'arez-Aubry and Jorma Louko.
\newblock {Onset and decay of the 1 + 1 Hawking-Unruh effect: what the
  derivative-coupling detector saw}.
\newblock {\em Class. Quant. Grav.}, 31(24):245007, 2014.

\bibitem{Louko:2014aba}
Jorma Louko.
\newblock {Unruh-DeWitt detector response across a Rindler firewall is finite}.
\newblock {\em JHEP}, 09:142, 2014.

\bibitem{Martin-Martinez:2014qda}
Eduardo Martin-Martinez and Jorma Louko.
\newblock {Particle detectors and the zero mode of a quantum field}.
\newblock {\em Phys. Rev. D}, 90(2):024015, 2014.

\bibitem{Tjoa:2022oxv}
Erickson Tjoa and Robert~B. Mann.
\newblock {Unruh-DeWitt detector in dimensionally-reduced static spherically
  symmetric spacetimes}.
\newblock {\em JHEP}, 03:014, 2022.

\bibitem{Ford:1993bw}
L.~H. Ford and Thomas~A. Roman.
\newblock {Motion of inertial observers through negative energy}.
\newblock {\em Phys. Rev. D}, 48:776--782, 1993.

\bibitem{DePaola:1996xw}
R.~De~Paola and N.~F. Svaiter.
\newblock {Radiative processes for Rindler and accelerating observers and the
  stress - tensor detector}.
\newblock 4 1996.

\bibitem{Sriramkumar:1994pb}
L.~Sriramkumar and T.~Padmanabhan.
\newblock {Response of finite time particle detectors in noninertial frames and
  curved space-time}.
\newblock {\em Class. Quant. Grav.}, 13:2061--2079, 1996.

\bibitem{Louko:2007mu}
Jorma Louko and Alejandro Satz.
\newblock {Transition rate of the Unruh-DeWitt detector in curved spacetime}.
\newblock {\em Class. Quant. Grav.}, 25:055012, 2008.

\bibitem{Schlicht:2003iy}
Sebastian Schlicht.
\newblock {Considerations on the Unruh effect: Causality and regularization}.
\newblock {\em Class. Quant. Grav.}, 21:4647--4660, 2004.

\bibitem{Louko:2006zv}
Jorma Louko and Alejandro Satz.
\newblock {How often does the Unruh-DeWitt detector click? Regularisation by a
  spatial profile}.
\newblock {\em Class. Quant. Grav.}, 23:6321--6344, 2006.

\bibitem{Lee:2012tvc}
Antony~R. Lee and Ivette Fuentes.
\newblock {Spatially extended Unruh-DeWitt detectors for relativistic quantum
  information}.
\newblock {\em Phys. Rev. D}, 89(8):085041, 2014.

\bibitem{Martin-Martinez:2020pss}
Eduardo Mart\'\i{}n-Mart\'\i{}nez, T.~Rick Perche, and Bruno de~S.~L.~Torres.
\newblock {General Relativistic Quantum Optics: Finite-size particle detector
  models in curved spacetimes}.
\newblock {\em Phys. Rev. D}, 101(4):045017, 2020.

\bibitem{Lin:2008jj}
Shih-Yuin Lin, Chung-Hsien Chou, and B.~L. Hu.
\newblock {Disentanglement of two harmonic oscillators in relativistic motion}.
\newblock {\em Phys. Rev. D}, 78:125025, 2008.

\bibitem{Lin:2008yz}
Shih-Yuin Lin and B.~L. Hu.
\newblock {Temporal and Spatial Dependence of Quantum Entanglement: Quantum
  'Nonlocality' in EPR from Field Theory Perspective}.
\newblock {\em Phys. Rev. D}, 79:085020, 2009.

\bibitem{Alsing:2012wf}
Paul~M. Alsing and Ivette Fuentes.
\newblock {Observer dependent entanglement}.
\newblock {\em Class. Quant. Grav.}, 29:224001, 2012.

\bibitem{Alhambra:2013uja}
\'Alvaro~M. Alhambra, Achim Kempf, and Eduardo Mart\'\i{}n-Mart\'\i{}nez.
\newblock {Casimir forces on atoms in optical cavities}.
\newblock {\em Phys. Rev. A}, 89(3):033835, 2014.

\bibitem{Martin-Martinez:2012ysv}
Eduardo Martin-Martinez, Miguel Montero, and Marco del Rey.
\newblock {Wavepacket detection with the Unruh-DeWitt model}.
\newblock {\em Phys. Rev. D}, 87(6):064038, 2013.

\bibitem{K:2021gns}
Hari K and Dawood Kothawala.
\newblock {Effect of tidal curvature on dynamics of accelerated probes}.
\newblock {\em Phys. Rev. D}, 104(6):064032, 2021.

\bibitem{Weinberg:2008zzc}
S.~Weinberg.
\newblock {\em Cosmology}.
\newblock Cosmology. Oxford University Press, Oxford, 2008.

\bibitem{Parker:1968mv}
L.~Parker.
\newblock {Particle creation in expanding universes}.
\newblock {\em Phys. Rev. Lett.}, 21:562--564, 1968.

\bibitem{Parker:1969au}
Leonard Parker.
\newblock {Quantized fields and particle creation in expanding universes. 1.}
\newblock {\em Phys. Rev.}, 183:1057--1068, 1969.

\bibitem{Parker:1971pt}
L.~Parker.
\newblock {Quantized fields and particle creation in expanding universes. 2.}
\newblock {\em Phys. Rev. D}, 3:346--356, 1971.
\newblock [Erratum: Phys.Rev.D 3, 2546--2546 (1971)].

\bibitem{Brandenberger:1984cz}
Robert~H. Brandenberger.
\newblock {Quantum Field Theory Methods and Inflationary Universe Models}.
\newblock {\em Rev. Mod. Phys.}, 57:1, 1985.

\bibitem{Parker:1999td}
Leonard Parker and Alpan Raval.
\newblock {Nonperturbative effects of vacuum energy on the recent expansion of
  the universe}.
\newblock {\em Phys. Rev. D}, 60:063512, 1999.
\newblock [Erratum: Phys.Rev.D 67, 029901 (2003)].

\bibitem{Parker:2009uva}
Leonard~E. Parker and D.~Toms.
\newblock {\em {Quantum Field Theory in Curved Spacetime}: {Quantized Field and
  Gravity}}.
\newblock Cambridge Monographs on Mathematical Physics. Cambridge University
  Press, 8 2009.

\bibitem{Lochan:2018pzs}
Kinjalk Lochan, Karthik Rajeev, Amit Vikram, and T.~Padmanabhan.
\newblock {Quantum correlators in Friedmann spacetimes: The omnipresent de
  Sitter spacetime and the invariant vacuum noise}.
\newblock {\em Phys. Rev. D}, 98(10):105015, 2018.

\bibitem{Garbrecht:2004du}
Bjorn Garbrecht and Tomislav Prokopec.
\newblock {Unruh response functions for scalar fields in de Sitter space}.
\newblock {\em Class. Quant. Grav.}, 21:4993--5004, 2004.

\bibitem{Garbrecht:2004ui}
Bjorn Garbrecht and Tomislav Prokopec.
\newblock {Energy density in expanding universes as seen by Unruh's detector}.
\newblock {\em Phys. Rev. D}, 70:083529, 2004.

\bibitem{Chakraborty:2019ltu}
Kushal Chakraborty and Bibhas~Ranjan Majhi.
\newblock {Detector response along null geodesics in black hole spacetimes and
  in a Friedmann-Lemaitre-Robertson-Walker Universe}.
\newblock {\em Phys. Rev. D}, 100(4):045004, 2019.

\bibitem{Hotta:2020pmq}
Masahiro Hotta, Achim Kempf, Eduardo Mart\'\i{}n-Mart\'\i{}nez, Takeshi
  Tomitsuka, and Koji Yamaguchi.
\newblock {Duality in the dynamics of Unruh-DeWitt detectors in conformally
  related spacetimes}.
\newblock {\em Phys. Rev. D}, 101(8):085017, 2020.

\bibitem{Ali:2020gij}
Md.~Sabir Ali, Sourav Bhattacharya, and Kinjalk Lochan.
\newblock {Unruh-DeWitt detector responses for complex scalar fields in de
  Sitter spacetime}.
\newblock {\em JHEP}, 03:220, 2021.

\bibitem{Conroy:2022rgp}
Aindri\'u Conroy.
\newblock {Unruh-DeWitt detectors in cosmological spacetimes}.
\newblock {\em Phys. Rev. D}, 105(12):123513, 2022.

\bibitem{Ford:1977in}
L.~H. Ford and L.~Parker.
\newblock {Infrared Divergences in a Class of Robertson-Walker Universes}.
\newblock {\em Phys. Rev. D}, 16:245--250, 1977.

\bibitem{Allen:1985ux}
Bruce Allen.
\newblock {Vacuum States in de Sitter Space}.
\newblock {\em Phys. Rev. D}, 32:3136, 1985.

\bibitem{Antoniadis:1985pj}
Ignatios Antoniadis, J.~Iliopoulos, and T.~N. Tomaras.
\newblock {Quantum Instability of De Sitter Space}.
\newblock {\em Phys. Rev. Lett.}, 56:1319, 1986.

\bibitem{Allen:1987tz}
Bruce Allen and Antoine Folacci.
\newblock {The Massless Minimally Coupled Scalar Field in De Sitter Space}.
\newblock {\em Phys. Rev. D}, 35:3771, 1987.

\bibitem{Polarski:1991ek}
D.~Polarski.
\newblock {Infrared divergences in de Sitter space}.
\newblock {\em Phys. Rev. D}, 43:1892--1895, 1991.

\bibitem{Miao:2010vs}
S.~P. Miao, N.~C. Tsamis, and R.~P. Woodard.
\newblock {De Sitter Breaking through Infrared Divergences}.
\newblock {\em J. Math. Phys.}, 51:072503, 2010.

\bibitem{Stargen:2021vtg}
D.~Jaffino Stargen and Kinjalk Lochan.
\newblock {Cavity Optimization for Unruh Effect at Small Accelerations}.
\newblock {\em Phys. Rev. Lett.}, 129(11):111303, 2022.

\bibitem{Kirsten:1993ug}
Klaus Kirsten and Jaume Garriga.
\newblock {Massless minimally coupled fields in de Sitter space: O(4) symmetric
  states versus de Sitter invariant vacuum}.
\newblock {\em Phys. Rev. D}, 48:567--577, 1993.

\bibitem{Dhanuka:2020yxp}
Ankit Dhanuka and Kinjalk Lochan.
\newblock {Stress energy correlator in de Sitter spacetime: Its conformal
  masking or growth in connected Friedmann universes}.
\newblock {\em Phys. Rev. D}, 102(8):085009, 2020.

\bibitem{Lochan:2022dht}
Kinjalk Lochan.
\newblock {Unequal time commutators in Friedmann universes: deterministic
  evolution of massless fields}.
\newblock {\em Gen. Rel. Grav.}, 54(9):100, 2022.

\bibitem{Page:2012fn}
Don~N. Page and Xing Wu.
\newblock {Massless Scalar Field Vacuum in de Sitter Spacetime}.
\newblock {\em JCAP}, 11:051, 2012.

\bibitem{Ford:1984hs}
L.~H. Ford.
\newblock {Quantum Instability of De Sitter Space-time}.
\newblock {\em Phys. Rev. D}, 31:710, 1985.

\bibitem{Tjoa:2020eqh}
Erickson Tjoa and Robert~B. Mann.
\newblock {Harvesting correlations in Schwarzschild and collapsing shell
  spacetimes}.
\newblock {\em JHEP}, 08:155, 2020.

\bibitem{Parker:1980kw}
L.~Parker.
\newblock {ONE ELECTRON ATOM AS A PROBE OF SPACE-TIME CURVATURE}.
\newblock {\em Phys. Rev. D}, 22:1922--1934, 1980.

\bibitem{Parker:1980hlc}
Leonard Parker.
\newblock {One-Electron Atom in Curved Space-Time}.
\newblock {\em Phys. Rev. Lett.}, 44(23):1559, 1980.

\bibitem{Fewster:2016ewy}
Christopher~J. Fewster, Benito~A. Ju\'arez-Aubry, and Jorma Louko.
\newblock {Waiting for Unruh}.
\newblock {\em Class. Quant. Grav.}, 33(16):165003, 2016.

\bibitem{Lochan:2014xja}
Kinjalk Lochan and T.~Padmanabhan.
\newblock {Inertial nonvacuum states viewed from the Rindler frame}.
\newblock {\em Phys. Rev. D}, 91(4):044002, 2015.

\bibitem{Kundu:2011sg}
Sandipan Kundu.
\newblock {Inflation with General Initial Conditions for Scalar Perturbations}.
\newblock {\em JCAP}, 02:005, 2012.

\bibitem{Fukuma:2013uxa}
Masafumi Fukuma, Yuho Sakatani, and Sotaro Sugishita.
\newblock {Master equation for the Unruh-DeWitt detector and the universal
  relaxation time in de Sitter space}.
\newblock {\em Phys. Rev. D}, 89(6):064024, 2014.

\bibitem{Tian:2013lna}
Zehua Tian and Jiliang Jing.
\newblock {Geometric phase of two-level atoms and thermal nature of de Sitter
  spacetime}.
\newblock {\em JHEP}, 04:109, 2013.

\bibitem{Tian:2014jha}
Zehua Tian and Jiliang Jing.
\newblock {Dynamics and quantum entanglement of two-level atoms in de Sitter
  spacetime}.
\newblock {\em Annals Phys.}, 350:1, 2014.

\bibitem{Vilenkin:1982wt}
Alexander Vilenkin and L.~H. Ford.
\newblock {Gravitational Effects upon Cosmological Phase Transitions}.
\newblock {\em Phys. Rev. D}, 26:1231, 1982.

\bibitem{Dolgov:1994cq}
A.~D. Dolgov, M.~B. Einhorn, and Valentin~I. Zakharov.
\newblock {On Infrared effects in de Sitter background}.
\newblock {\em Phys. Rev. D}, 52:717--722, 1995.

\bibitem{Marolf:2010zp}
Donald Marolf and Ian~A. Morrison.
\newblock {The IR stability of de Sitter: Loop corrections to scalar
  propagators}.
\newblock {\em Phys. Rev. D}, 82:105032, 2010.

\bibitem{Padmanabhan:1987rq}
T.~Padmanabhan and T.~P. Singh.
\newblock {Response of an Accelerated Detector Coupled to the Stress - Energy
  Tensor}.
\newblock {\em Class. Quant. Grav.}, 4:1397--1407, 1987.

\bibitem{Perez-Nadal:2009jcz}
Guillem Perez-Nadal, Albert Roura, and Enric Verdaguer.
\newblock {Stress tensor fluctuations in de Sitter spacetime}.
\newblock {\em JCAP}, 05:036, 2010.

\bibitem{Hu:2008rga}
B.~L. Hu and E.~Verdaguer.
\newblock {Stochastic Gravity: Theory and Applications}.
\newblock {\em Living Rev. Rel.}, 11:3, 2008.

\bibitem{Dai:2015rda}
Liang Dai, Enrico Pajer, and Fabian Schmidt.
\newblock {Conformal Fermi Coordinates}.
\newblock {\em JCAP}, 11:043, 2015.

\bibitem{Mukhanov:1990me}
Viatcheslav~F. Mukhanov, H.~A. Feldman, and Robert~H. Brandenberger.
\newblock {Theory of cosmological perturbations. Part 1. Classical
  perturbations. Part 2. Quantum theory of perturbations. Part 3. Extensions}.
\newblock {\em Phys. Rept.}, 215:203--333, 1992.

\bibitem{Bartolo:2004if}
N.~Bartolo, E.~Komatsu, Sabino Matarrese, and A.~Riotto.
\newblock {Non-Gaussianity from inflation: Theory and observations}.
\newblock {\em Phys. Rept.}, 402:103--266, 2004.

\bibitem{Poisson:2009pwt}
Eric Poisson.
\newblock {\em {A Relativist's Toolkit: The Mathematics of Black-Hole
  Mechanics}}.
\newblock Cambridge University Press, 12 2009.

\bibitem{Griffiths1}
David~J. Griffiths and Darrell~F. Schroeter.
\newblock {\em Introduction to Quantum Mechanics}.
\newblock Cambridge University Press, 3 edition, 2018.

\end{thebibliography}

%
%

%

\end{document}